%% file: main.tex
\documentclass[sigplan,10pt]{acmart}
\renewcommand\footnotetextcopyrightpermission[1]{}
\AtBeginDocument{%
  }

\usepackage{algorithm}
\usepackage{algpseudocodex}
\usepackage{enumitem}
\usepackage{tikz}
\usetikzlibrary{positioning, arrows.meta, shapes.geometric}
\usepackage{adjustbox}
\usepackage[capitalise,noabbrev]{cleveref}
\usepackage{changepage}
\usepackage{makecell}
\usepackage{multirow}
\usepackage{tabularx}
\usepackage{array}

\input{macros}

\begin{document}

\title{A Kubernetes Scheduler Plugin for Cluster-Wide Placement Optimisation}

\author{Henrik Daniel Christensen}
\affiliation{%
  \institution{TV 2 Danmark A/S}
  \country{Denmark}
}
\email{hech@tv2.dk}

\author{Saverio Giallorenzo}
\affiliation{%
  \institution{Università di Bologna and INRIA}
  \country{Italy and France}
}
\email{saverio.giallorenzo@unibo.it}

\author{Jacopo Mauro}
\affiliation{%
  \institution{University of Southern Denmark}
  \country{Denmark}
}
\email{mauro@imada.sdu.dk}

\input{sections/00_abstract.tex}



\received{20 February 2007}
\received[revised]{12 March 2009}
\received[accepted]{5 June 2009}

\maketitle
\pagestyle{plain}

\algrenewcommand{\algorithmiccomment}[1]{%
  \hfill \textit{\color{gray}\(\triangleright\) #1}}
\algrenewcommand\algorithmicindent{0.8em}

\newcommand{\fourvec}[4]{\ensuremath{\langle #1,\,#2,\,#3,\,#4\rangle}}

\input{sections/01_introduction}
\input{sections/02_preliminaries}
\input{sections/03_system_design}
\input{sections/04_evaluation}

\input{sections/05_related_work}
\input{sections/06_conclusions}
\bibliographystyle{ACM-Reference-Format}
\bibliography{bibliography}

\appendix

\newpage
\input{appendix/algorithms}

\newpage
\input{appendix/public_trace_analysis}

\newpage
\input{appendix/optimisation_and_solver}

\newpage
\input{appendix/results}

\end{document}

%% file: macros.tex
\usepackage[
  disable,     %
  textwidth=\marginparwidth,
  colorinlistoftodos,
  prependcaption,
  linecolor=green,backgroundcolor=green!25,bordercolor=green
  ]{todonotes}
\makeatletter
\if@todonotes@disabled
\else %
\newlength{\increase}
\paperwidth=\dimexpr \paperwidth + \increase\relax
\oddsidemargin=\dimexpr\oddsidemargin + .5\increase\relax
\evensidemargin=\dimexpr\evensidemargin + .6\increase\relax
\marginparwidth=\dimexpr \marginparwidth + .4\increase\relax
\let\tmptitle\title
\renewcommand{\title}[1]{\tmptitle{#1{\tiny\normalfont \makebox[10em]{\colorbox{red!20}{disable todos}}}}}
\fi
\makeatother

\usepackage{xspace}
\newcommand{\tool}{\texttt{OPSche}\xspace}

%% file: sections/00_abstract.tex
\begin{abstract}

The default scheduler of Kubernetes, the state-of-the-art container
orchestrator, uses fast, local placement decisions. Unfortunately, this design
leads to resource fragmentation, reduced cluster usage, and overprovisioning.
External solvers can compute global placement plans, but enforcing these plans
in upstream clusters is hard. Kubernetes provides no native cross-node
preemption, uncoordinated concurrent scheduling leads to inconsistencies, and
replacing the default scheduler would sever deployments from upstream cycles.

We present \tool, an open-source Kubernetes Scheduling Framework plugin where
external solvers can drive cluster-wide placement decisions in concert with the
default scheduler. \tool atomically validates and enforces solver-produced plans
through coordinated framework hooks and supports three trigger modes:
scheduling-failure, periodic, and stable-queue -- resp.\@ triggered when a
workload cannot be placed, at fixed time intervals, when the set of pending
workloads stabilises. Each mode has a blocking variant for a finer tuning of
placement quality, latency, and disruption.

We pair \tool with a constraint-based optimisation solver, showing its
feasibility across a broad set of cluster configurations and reporting
improvements of resource usage by up to 3.0\% and scheduling latency by more
than a second.

\end{abstract}

%% file: sections/01_introduction.tex
\section{Introduction}
\label{sec:introduction}

Kubernetes is the state-of-the-art for container orchestration, managing
workloads across clusters that range from a handful of nodes in small
enterprises to thousands in hyperscale deployments~\cite{cncf2024survey}. At the
heart of Kubernetes sits its scheduler: the component responsible for deciding
which node should run each pending pod -- the smallest deployable unit in
Kubernetes that encapsulates one or more containers.

\paragraph{The problem: local vs.\ global decisions.}
A problem that characterise Kubernetes default scheduler is that it relies on
local, greedy decisions where it evaluates each pod independently, scores the
available nodes, and commits to the highest-scoring one, without any
coordination across the set of pods under scheduling. This local strategy is
fast and simple, but it leaves performance on the table: fragmentation
accumulates silently, nodes end up partially filled, and pods get stuck pending,
whereas a smarter, cluster-wide rearrangement could accommodate
them~\cite{CAD18}.

To get a sense of the problem, consider a cluster where two nodes are each
half-full with pods of different sizes. A new pod arrives that fits on neither
node individually, yet it would fit if one pod moved to the other node. In such
situations, the default scheduler would see a scheduling failure and give up,
whereas a cluster-wide view would reveal the fix.
This inefficiency is reflected in practice: industry surveys report that in 2025
the gap between provisioned and requested resources averaged 40\% for CPU and
57\% for memory~\cite{castai2025k8sCostBenchmark} with 99.94\% of all the
clusters overprovisioned on CPUs.

Techniques to improve/optimise global placement plans abound (complete search
methods, heuristics, etc.), however, the obstacle is not (solely) in performance
but in affordance. The Kubernetes scheduler does not natively implement a global scheduling logic and replacing it with a custom one would sever deployments from upstream release cycles.

\paragraph{The challenge: atomically enforcing a plan.}
Enforcing a global scheduling plan on Kubernetes scheduler has two main
obstables. First, Kubernetes provides no native mechanism for \emph{cross-node
preemption} -- its \texttt{DefaultPreemption} routine frees capacity on a single
candidate node, and cannot coordinate evictions across multiple nodes to realise
a cluster-wide rearrangement. Second, the default scheduler runs multiple
workers concurrently, each independently dequeuing and placing pods. Without
explicit coordination, concurrent workers may consume the capacity that the plan
relies on, rendering the plan obsolete before its enforcement.

One could sidestep both issues by replacing the default scheduler
entirely with a custom implementation. However, doing so would sever the
deployment from the upstream Kubernetes release cycle -- one would need to
manually port every bug fix, security patch, and feature improvement shipped in
the official scheduler, imposing a significant and open-ended maintenance
burden. Thus, we deem that the right design goal is to {extend the default
scheduler} -- preserving full compatibility with upstream Kubernetes while
adding the coordination layer that global optimisation requires.

\paragraph{The proposed solution: \tool} We present \tool, a Kubernetes
scheduler plugin that enables an external solver to drive cluster-wide placement
decisions in concert with the default scheduler. We build \tool entirely on top
of the \emph{Kubernetes Scheduling Framework} -- the official extension
mechanism introduced in Kubernetes v1.15 -- so that it requires no modifications
to any upstream scheduler code.

The core insight behind \tool is that, by using five hooks that the Scheduling
Framework provides, one can enforce arbitrary cluster-wide placement plans
atomically and safely preserving concurrent scheduling. Essentially, \tool works
by taking a snapshot of the current cluster state, invoking a solver that
computes a deployment plan, and validating the returned plan against a fresh
snapshot to prevent the application of stale plans on drifted states. If the
plan is valid, \tool enforces it in three phases: \emph{eviction}, where it
issues eviction requests for all pods marked for removal or relocation;
\emph{activation}, where it re-activates only the pods that the plan selects for
placement, holding all other inbound pods back; and \emph{realisation}, where it
steers each pod to its designated node by constraining placement to exactly one
feasible node per pod. \tool prevents concurrent over-commitment of nodes by
tracking per-node resource budgets.

To support a wide set of dynamics and operational constraints found in
real-world clusters, \tool provides three trigger modes: a
\emph{scheduling-failure} mode that reacts immediately when a pod becomes
unschedulable, and two {background} modes -- \emph{periodic} and
\emph{stable-queue} -- that run the solver proactively, either at fixed
intervals or only once the pending-pod set has stabilised. Cluster-wide
placement reduces to an NP-hard combinatorial optimisation problem and solver
execution may incur non-negligible latency under high contention. For this
reason, each mode has two variants: a \emph{blocking}, which pauses scheduling
during solver execution, and a \emph{non-blocking}, which allows normal
scheduling to proceed in parallel. This plethora of modes lets operators tune
the trade-off between placement quality, scheduling latency, and operational
disruption to match their workload regime.

We evaluate \tool on emulated Kubernetes clusters built with \emph{Kubernetes
WithOut Kubelet}~\cite{kubernetes_developers_kwok_2025}, driven by synthetic
traces parameterised from the Alibaba
GPU~\cite{alibaba_alibabacluster-trace-gpu-v2025_2025} and Google Cluster
Data~\cite{google_developers_googleclusterdata2019_2025} production traces. The
main experimental result is a qualitative one, i.e., the fact that \tool
correctly implements global scheduling as a Kubernetes scheduler plugin.
Quantitatively, testing our plugin across 16- and 32-node clusters, two priority
configurations, and three workload arrival regimes, we found that \tool
consistently improves effective resource usage by 1.2--3.0\% over the default
scheduler (single priority) and by 0.2--1.4\% under four priority levels, while
the periodic and stable-queue modes also reduce scheduling latency by more than a second.
The non-blocking variant matches the blocking
variant's resource efficiency while imposing less interference on regular
scheduling, emerging as the recommended default.

\paragraph{Paper Outline.}
We review Kubernetes scheduling and the Scheduling Framework in
\cref{sec:preliminaries}. Then, we present the design of \tool in
\cref{sec:system_design}, followed by the description of our evaluation setup
and results in \cref{sec:evaluation}. We conclude discussing related work in
\cref{sec:related_work} and drawing final remarks and future directions in
\cref{sec:conclusions}.

%% file: sections/02_preliminaries.tex
\section{Preliminaries}
\label{sec:preliminaries}

We provide the background needed to understand the design and implementation of
our scheduler plugin. We review key aspects of Kubernetes' architecture and
default scheduler, and introduce the Kubernetes Scheduling Framework.

\subsection{Kubernetes Architecture and Core Concepts}

A Kubernetes cluster consists of a control plane (including the API server) and
worker nodes. The smallest deployable unit is a pod, which groups one or more
containers, each enclosing one or more applications and runtime environments.

Workload controllers (e.g., \texttt{Deployments}, \texttt{ReplicaSets})
typically create and maintain pods, keep a desired replica count -- each replica
corresponding to a pod instance -- and manage pod lifecycle events such as pod
replacement.

Each container can specify CPU and memory requests, which determine the
resources reserved for the pod and guide the scheduler. Each worker node exposes
a finite capacity, of which the allocatable portion is available to pods.

\paragraph{Kubernetes Default Scheduler}

The {default scheduler} in Kubernetes assigns nodes to pending pods through
a per-pod scheduling cycle. Pending pods are managed in a {queueing
system}\footnote{The system has three queues: \emph{active} (ready for
scheduling), \emph{unschedulable} (failed scheduling), and
\emph{backoff} (temporary hold after repeated failures).}, and are processed in
parallel by multiple {scheduler workers}.
For each pod, a worker filters nodes based on {feasibility constraints}
(e.g., sufficient resources, node selectors), scores the
feasible nodes, and selects the highest-scoring node. As a result, the scheduler
makes {local}, {heuristic} decisions based on the current cluster
state, without explicit coordination across multiple pending pods.

\paragraph{Kubernetes Scheduling Framework}

One can customise Kubernetes scheduling in multiple ways, including running a
separate scheduler and modifying the scheduler code. However, these approaches
typically incur significant engineering and maintenance overhead. In response,
Kubernetes v1.15 introduced the \emph{Scheduling Framework}, which allows users
to inject custom logic into the default scheduler without having to implement
and maintain an entire scheduler from scratch.

The Scheduling Framework structures scheduling into extension points -- also
called \emph{hooks} -- that execute at different stages of the scheduling
cycle~\cite{kubernetes_developers_scheduling_2024} and that plugins can attach
to for implementing custom scheduling logic cleanly integrated with the default
scheduler.

In particular, the default scheduler already includes a built-in preemption
mechanism, \texttt{DefaultPreemption},
invoked when no feasible node is found for a pod to evict lower-priority pods.
However, this mechanism has a {single-node} scope: it selects a candidate node
and preempts pods therein until it frees enough resources. Thus, Kubernetes
provides no native support for {multi-node} (cross-node) preemption, i.e.,
eviction strategies that coordinate pod removals across multiple nodes to
realise a cluster-wide placement decision.

%% file: sections/03_system_design.tex
\section{System Design}
\label{sec:system_design}
We present the design of \tool, our {optimisation plugin} for the
Kubernetes default scheduler. In particular, we discuss two key design dimensions behind
\tool: (1) when to trigger optimisation and whether to block concurrent
scheduling during solver execution, and (2) how to enforce placement plans
through the hooks while coordinating multiple parallel
scheduler workers that independently schedule pods. Thanks to the combination of
the presented design choices, \tool implements optimal pod scheduling for
Kubernetes as a plugin that one can install alongside a vanilla Kubernetes
deployment, leaving its core components unchanged.

\subsection{\tool Workflow}
\label{sec:optimisation_execution}

\begin{figure}[t]
    \centering
    \includegraphics[width=.8\columnwidth, clip, trim=0.75cm 0.7cm 0.6cm 0.5cm]{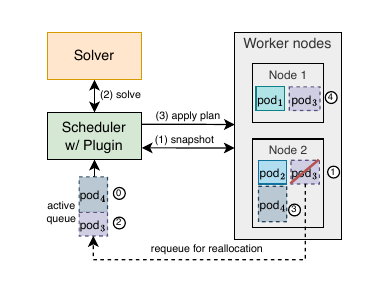}
    \caption{\label{fig:plugin-design}Example of \tool's workflow: a pod (pod\(_3\)) is moved from Node 2 to Node 1 to place a pending pod
    (pod\(_4\)).}
\end{figure}

Conceptually, the workflow of \tool is simple, it (1) snapshots the current
cluster state, (2) asks for an improved plan from a solver that may relocate or
evict pods across nodes, and (3) enforces this plan by coordinating how the
scheduler admits new pods and re-schedules pods marked for relocation. We
visualise \tool's workflow in \cref{fig:plugin-design} through an example of
cluster-wide optimisation via cross-node reallocation. Initially, Node 1 hosts
\textsf{pod}\(_{1}\) and Node 2 hosts \textsf{pod}\(_{2}\) and
\textsf{pod}\(_{3}\). When \textsf{pod}\(_{4}\) arrives \textcircled{0}, only
Node 2 could have sufficient capacity to host it, but the presence of both
\textsf{pod}\(_{2}\) and \textsf{pod}\(_{3}\) prevents its deployment. The
workflow proceeds as follows: (1) \tool captures the cluster state; (2) the
solver determines that \textsf{pod}\(_{4}\) can be placed on Node 2 if
\textsf{pod}\(_{3}\) relocates to Node 1; (3) \tool applies the plan by evicting
\textsf{pod}\(_{3}\) \textcircled{1} -- which re-enters the active queue
\textcircled{2} -- and pinning \textsf{pod}\(_{4}\) to Node 2 \textcircled{3}
and \textsf{pod}\(_{3}\) to Node 1 \textcircled{4}.

\paragraph{Snapshot construction}

When \tool triggers an optimisation run, it first constructs a snapshot of the
current cluster state from shared informer caches by extracting the relevant node and
pod state (both running and pending). Informers are event-driven clients that
maintain an up-to-date in-memory cache streamed from the Kubernetes API server.
The scheduler consults these caches
 when selecting nodes. From the extracted state, \tool
passes the snapshot to an external optimisation solver. \tool is
solver-agnostic. Any solver able to produce a placement plan that satisfies the
Kubernetes placement constraints would work, whether it be based on heuristics
or a complete search procedure (e.g., constraint programming, SAT, SMT or MIP).
The solver should be interruptible at any time and, if possible, return the best
feasible plan found so far; otherwise no plan.

\paragraph{Plan structure and validation}

The plan returned by the solver describes three kinds of actions: (i) placements
for pending pods, (ii) movements of already running pods between nodes, and
(iii) evictions of pods that must be removed to free capacity. Before accepting
the plan, \tool validates that the former is still applicable, since the cluster
state may have drifted after taking the {snapshot}.

The validation step checks the state of a fresh snapshot against the plan's
preconditions to ensure that a) pods to be moved still exist and reside on their
initial node; b) pods marked for placement still exist and are pending; and c)
each node is still usable. \tool then re-evaluates the plan's resource
feasibility w.r.t.\@ the capacity it would free (evictions) and the capacity it
would consume (placements and moves) to ensure every node stays within its
allocatable CPU/memory limits. If any check fails, the plan is discarded, which
avoids overfilling nodes or acting on pods/nodes that have moved, terminated, or
otherwise changed.

\paragraph{Plan enforcement}

If a plan passes validation, \tool enforces it. \tool has at most one active
plan at any time, used to store and track the pods to be placed and moved and
that are held back from re-entering the scheduling queue.

Plan enforcement goes through three phases: \emph{eviction}, \emph{activation},
and \emph{realisation}.

At eviction, \tool issues eviction requests for every
pod that the current plan marks for move/removal. Evictions are irrevocable:
after issuing the first eviction, one cannot restore the original cluster state.
To avoid scheduling pods before capacity is available, \tool waits for the
required evictions to complete before proceeding. Concretely, \tool polls for
pods to disappear at a fixed interval (250\,ms by default).

During activation, \tool re-activates the set of pending pods that the plan
selects for placement, including pods that were evicted and are expected to
reappear as pending (e.g., via their controllers). Meanwhile, inbound pods that
are outside the plan are held back to prevent them from consuming capacity that
the placement plan relies on.

At realisation, admission and node selection are constrained to enforce the
plan: selected pods are admitted and pinned to their designated destination
nodes. A watcher monitors progress at a fixed interval (250\,ms by default) and
declares the plan complete once it observes the intended placements, otherwise,
after a set timeout (4\,s by default), it marks the plan as failed. Then, \tool
clears the active plan and releases any pods that were held back during
execution and the scheduler resumes normal operation.

\subsection{Operational Modes}
\label{sec:operational_modes}

\tool delegates finding better cluster-wide placements to an external
optimisation solver.
There are two orthogonal axes that determine solver invocations: (1) when to run
the solver, and (2) whether solver execution blocks the scheduler from
processing inbound pods (\emph{blocking} vs. \emph{non-blocking}).

\paragraph{When to run the solver}

Regarding the first axis, \tool supports a \emph{scheduling-failure} mode, which
triggers a solver run immediately if the default scheduler fails to find a node
for the pod to be placed, and two background modes, \emph{periodic} and
\emph{stable-queue}. The periodic mode runs at fixed intervals, whereas the
stable-queue mode triggers only after the queue of pending pods has remained
stable for a idle time window.

In \emph{scheduling-failure} mode, each scheduling failure triggers a solver run
that attempts to schedule the failing pod and, if a new plan is found,
rearranges or evicts other pods. The advantage of this mode is that it is
responsive (to failures), but frequent pod arrivals may cause bursts of solver
invocations adding latency to other pods.

In \emph{periodic} mode, \tool runs the solver at fixed intervals (by default
8\,s). This mode amortises solver costs, but pods that become unschedulable just
after the triggering of the solver would have to wait until the next run. To
avoid wasting work in stable periods, \tool records cluster states it has
already solved and, if the same state reappears, the solver invocation is
skipped.

In \emph{stable-queue} mode, \tool runs the solver only after the set of pending
pods has remained unchanged for a configurable idle window (by default 8\,s).
Stability is checked by polling the pending set at a configurable interval
(250\,ms by default). Moreover, if the pending set changes during solver
execution (e.g., a new pod arrives in the queue), the current solver run is
cancelled, and \tool returns to waiting for a stable window before starting a
new run. By cancelling an ongoing solver run whenever the pending set changes,
\tool reduces the likelihood of producing plans that are stale by the time they
are enforced. On the downside, frequent pod arrivals prevent the occurrence of
the idle window, leading to infrequent optimisation.

\paragraph{Blocking vs non-blocking plan computation}

Regarding the second axis, \tool supports both \emph{blocking} and
\emph{non-blocking} modes, which determine whether the scheduler can process
inbound pods during solver execution.

In \emph{blocking} mode, \tool acquires a plan lock before invoking the solver
and holds it until a plan has been enforced or discarded. While \tool has the
lock, both new pods and parallel attempts to start a new plan are put on hold.

In \emph{non-blocking} mode, \tool runs the solver without taking the lock;
thus, inbound pods can continue to be scheduled by the default scheduler during solver execution. If the solver returns an improving plan, \tool checks the
plan against the fresh snapshot and, if the plan is applicable (i.e., no
incoming scheduling requests have made the plan outdated), it is enforced.
Therefore, the non-blocking mode allows inbound pods to be scheduled at the
cost of occasionally discarding plans due to state drift.\footnote{Note that to
verify that the plan is still applicable, \tool must re-validate it against a
fresh snapshot, which requires acquiring the plan lock. This behaviour does not
present an issue, considering that the lock is held only briefly during
validation and before enforcement and not for the entire duration of the
optimisation run, as in blocking mode.}

In any combination of modes, once a plan is accepted, its enforcement proceeds
under the plan lock until the plan is realised or times out.

\subsection{Integrating \tool into the Kubernetes Scheduler}

\begin{figure}[t]
    \centering
    \includegraphics[width=\columnwidth]{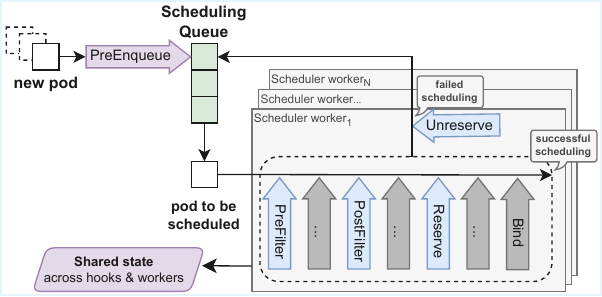}
    \caption{\label{fig:pluginArchitecture}\tool's interations with
    Kubernetes Scheduling Framework's hooks to admit, filter,
    reserve, and bind pods.}
\end{figure}

\tool implements scheduling optimisation by interacting with some of the hooks
provided by the Kubernetes Scheduling Framework as visualised in \cref{fig:pluginArchitecture}.
In the
following, we briefly comment on how \tool interacts with those hooks.

\paragraph{Plan computation}

To support the different trigger modes, \tool invokes the solver either (i) when
the scheduler worker reaches the \emph{PostFilter} hook after a failed
scheduling attempt (\emph{scheduling-failure} mode), or (ii) from a background
routine in the \emph{periodic} and \emph{stable-queue} modes.

\paragraph{Plan enforcement}

The Kubernetes scheduler processes pods in parallel using multiple scheduler
workers that each dequeue a pod and execute a full scheduling cycle
independently. \tool uses a small shared state that stores the current active
plan (if any) to coordinate the plan enforcement. \tool enforces the application
of an active plan by constraining scheduler workers to mark only the
plan-designated node as feasible for a given pod. The shared state ensures that,
at any time, there exists at most one optimisation run and one active plan and
that parallel workers apply consistent reservations and do not over-commit
nodes. To implement this behaviour, \tool extends the \emph{PreEnqueue},
\emph{PreFilter}, \emph{Post-Filter}, \emph{Reserve}, and \emph{Unreserve}
hooks. \cref{fig:pluginArchitecture} shows the position of these hooks in the
Scheduling Framework and how they interact with an active plan.

We briefly comment on the activation sequence of these hooks w.r.t.\@ pod
scheduling and how \tool makes use of them. When a new pod arrives, the worker
first calls the \emph{PreEnqueue} hook to decide whether the pod may enter the
scheduling queues. \tool uses this hook to temporarily hold back pods that are
not part of the plan, so they do not interfere with its execution. When a worker
is ready to schedule the next pod, it dequeues the pod from its queue and
proceeds to run the scheduling routine, passing through the remaining hooks. The
worker calls the \emph{PreFilter} hook to compute per-node feasibility. \tool
uses this hook to exclude nodes that are not part of the active optimisation
plan, effectively placing pods on the nodes chosen by the plan. If no feasible
node is found, the worker invokes the \emph{PostFilter} hook, from which \tool's
\emph{scheduling-failure} mode invokes the optimiser. To ensure that parallel
workers do not over-commit nodes, the worker invokes the \emph{Reserve} hook to
tentatively reserve capacity for the pod on its candidate node before binding
it. \tool leverages the same mechanism when enforcing a plan: it maintains
per-node budgets that track how much of each workload can still be placed on a
node. When a worker reserves a pod, \tool decrements the corresponding budget;
if the scheduling attempt later fails, the \emph{Unreserve} hook is invoked to
roll back the reservation and refund the budget.\footnote{To reliably relocate
pods via eviction, \tool identifies pods by their workload owner -- i.e., the
managing controller (e.g., a ReplicaSet) -- rather than by concrete pod
identities such as names or UIDs. The reason for this choice is that pod
identities are ephemeral: once a pod is evicted, its controller may recreate it
with a new identity. Accordingly, when a plan requires moving a pod, \tool
records the workload owner, reserves capacity for that workload on the target
node, and then evicts the pod, delegating the actual recreation to the
controller.}

\paragraph{Checking plan completion}

To detect when a plan has completed, \tool runs a background watcher that
periodically compares the current cluster state against the active
plan.\footnote{Kubernetes experts might wonder why we do not use the
\emph{PostBind} hook -- which runs immediately after the scheduler commits a
binding to the API server -- to obtain the completion signal. The reason is that
pods that are part of a plan can be deleted, recreated, or scaled down by their
controllers before reaching the hook, which implies that it can happen that the
``last'' pod, which would signal a plan's completion, might disappear,
preventing \tool from reliably inferring the plan's status.} The watcher checks
completion w.r.t.\@ all pods relevant to the plan. The completion check treats
pods differently depending on their category.

Pods in the \emph{identity-stable} category are standalone pods, created
directly as \texttt{Pod} objects without being managed by a higher-level
controller. The identity (name/UID) of these pods is stable during their
lifetime and the plan can refer to their concrete pod instance. In this case,
the watcher treats a planned placement as satisfied once it observes that the
pod is either bound to its designated node (i.e., \texttt{pod.spec.nodeName}
matches the planned destination), or terminating/absent (in which case, the
concrete instance no longer exists and plan enforcement cannot proceed).
Pods in the \emph{identity-unstable} category are controller-managed pods,
created and maintained by a controller (e.g., \texttt{ReplicaSet},
\texttt{Deployment}). Controllers may delete and recreate pods at any time
(e.g., after eviction), typically producing new pod names and UIDs.
Consequently, the plan cannot reliably track individual pod identities. Instead,
the watcher tracks progress at the workload level (via the workload owner
reference) and reasons in terms of quotas: a planned workload placement is
satisfied once the workload has no remaining pending pods attributed to the plan, or the workload has no pods because
it was scaled down or deleted.

When the plan has completed, the watcher clears the active plan and releases all
pods held back during the plan's execution, allowing the scheduler to resume
regular scheduling. To avoid blocking the system in pathological cases where
pods remain pending indefinitely (e.g., due to constraints extrinsic to our
model, external controllers, or incompatible node conditions) there is a plan
enforcement timeout that aborts enforcement if the plan has not completed within
a configurable time window (4\,s by default). When the timeout expires, the plan
is discarded, any held-back pods are released, and regular scheduling resumes.

%% file: sections/04_evaluation.tex
\section{Evaluation}
\label{sec:evaluation}

This section describes our evaluation of \tool across its operating modes and
compares it with the Kubernetes default scheduler, quantifying the impact on
resource usage, scheduling latency, pod deletions, and optimisation overhead.

\subsection{Methodology}
\label{sec:evaluation-methodology}

We evaluate \tool using a lightweight, trace-driven framework that runs on a
single machine while simulating cluster dynamics over extended time horizons.
The framework comprises three main elements: cluster emulation, trace
generation, and trace replay.

\paragraph{Cluster Emulation}

To simulate Kubernetes clusters, we use Kubernetes WithOut
Kubelet (KWOK)~\cite{kubernetes_developers_kwok_2025}, an open-source emulator
that instantiates a Kubernetes control plane with virtual nodes. KWOK faithfully
emulates Kubernetes scheduling logic while avoiding the resource overhead of
executing containers. Each virtual node exposes CPU and memory capacity and pods
transition through standard Kubernetes scheduling phases and consume simulated
capacity based on their resource requests. Because KWOK advances the system
state in real time, the duration of an experiment directly corresponds to
wall-clock time and enables us to run experiments using production scheduler
binaries (with and without \tool) while achieving reproducible results.

\paragraph{Trace Generation}

At the time of writing, there is no publicly available benchmark for Kubernetes
scheduling and preemption policies. For this reason, we develop a \emph{trace
generator} that produces workload arrivals and removals, designed to reflect
real-world workloads in a cluster. Each trace specifies the number of nodes in
the cluster and a time-ordered sequence of workload events (arrivals and
removals) over a fixed simulation horizon. We instantiate each workload as a
\texttt{ReplicaSet} that manages one or more pod replicas. The trace generator
can produce a wide range of workload patterns by configuring the distributions
used to sample workload attributes, including inter-arrival times, lifetimes,
priority levels, replica counts, and per-pod resource requests.

\begin{table}[t]
\centering
\caption{Trace generator parameters and values used in the experiments.}
\label{tab:trace-params}
\begin{adjustbox}{width=\columnwidth}
\begin{tabular}{ll}
\toprule
\textbf{Trace parameter} & \textbf{Values used} \\
\midrule
Nodes in cluster & $\{16,32\}$ \\
Simulation time & $2\,\mathrm{h}$ \\
Target resource usage & $90\%$ \\
Workload inter-arrival time: mean & $\{4,8,16\}\,\mathrm{s}$ \\
Workload inter-arrival time: min & 0.1 ms \\
Workload inter-arrival time: max & $16 \times \text{mean inter-arrival time}$ \\
Workload lifetime: min & $2\,\mathrm{s}$\\
Workload lifetime: max & $24\,\mathrm{h}$ \\
Workload priority & $\{1\}$ or $\{1,2,3,4\}$ (geom. ratio $0.8$) \\
Replicas per \texttt{ReplicaSet} & $\{1,2,3\}$ (geom. ratio $0.8$) \\
Per-pod resource requests: mean & $0.125$ (fraction of a node) \\
Per-pod resource requests: min & $0.05$ (fraction of a node) \\
Per-pod resource requests: max & $0.5$ (fraction of a node) \\
\bottomrule
\end{tabular}%
\end{adjustbox}
\end{table}

To generate realistic traces, we first analyse publicly available
traces~\cite{liu_public_2025}, particularly the Alibaba
GPU~\cite{alibaba_alibabacluster-trace-gpu-v2025_2025} and Google Cluster Data
2019~\cite{google_developers_googleclusterdata2019_2025} traces, to parameterise
the distributions that underlie these attributes.\footnote{We provide an analysis of the traces in
Supplementary Material, \cref{appendix:public-trace-analysis}.} These
traces exhibit heavy-tailed inter-arrival times, lifetimes, and resource
requests, which we model using truncated Pareto (Type I) distributions specified
by a target mean and min/max bounds. We model priorities and replica counts
using truncated geometric distributions, restricting replica counts and number
of priority levels to a bounded range. The list of parameters of the trace
generation process, including the parameters of the distributions, is available
in \cref{tab:trace-params}.

To stress-test the scheduler, we generate traces with a high degree of
\emph{resource contention} -- i.e., situations where competing pods share
limited cluster resources (e.g., due to high demand, fragmentation, or placement
constraints). 
To this end, we first define the notion of \emph{effective usage} at each time
unit as the highest of normalised CPU and memory requests, and the
\emph{time-averaged effective usage} as the mean of effective usage over the
simulation horizon. Thus, to achieve high resource contention, the trace
generator calibrates the workload lifetime distribution so that the resulting
trace reaches a target \emph{time-averaged effective usage} through the repeated
adjustment of the mean lifetime for each trace configuration.
We obtain those traces by finding an initial lifetime estimate which we then
refine iteratively. Practically, given the number of nodes, the mean
inter-arrival time, and the configured distributions for resource requests and
replica counts, we obtain an initial mean lifetime estimate via Little's
Law~\cite{little_proof_1961}. Starting from the initial lifetime estimate, we
generate a full trace realisation and measure its time-averaged effective usage.
If the measured usage deviates from the target, we scale the mean lifetime
proportionally and repeat the procedure. The process stops once the relative
error is within $1\%$.

To avoid cold-starts (where there is initially little to no contention), each
generated trace includes an initial state containing workloads that are already
running at the beginning of the simulation, such that their aggregate effective
requested load approximately matches the target usage. Each initial workload is
created with start time $0$ and a sampled remaining lifetime. One could follow a
na\"ive approach and sample the remaining lifetime directly from the truncated
Pareto distribution $f(x)$ used for newly arriving workloads. However, this
practice would systematically underrepresent long-lived workloads in the initial
state since -- intuitively, at any observation point, workloads with longer
lifetimes are more likely to be present than short-lived ones. Following the
standard correction for this bias~\cite{qin_examples_2017}, we shift the
distribution by sampling initial workloads from the length-biased distribution
$f^\star(x) \propto x \cdot f(x)$, that for a Pareto distribution with $f(x)
\propto x^{-(\alpha+1)}$ yields $f^\star(x) \propto x^{-\alpha}$. After sampling
a total lifetime $X^\star$ from $f^\star$, we assign each initial workload a
remaining lifetime $\sim \mathrm{Uniform}(0, X^\star)$, which corresponds to
observing the workload at a random point during its execution. We populate the
initial state by repeatedly sampling workloads (remaining lifetime, replica
count, priority, and CPU/memory requests) until the effective requested load
reaches the target.

Finally, we generate the newly arriving workloads (workloads after the initial
state) by sampling workloads sequentially until the simulation horizon is
covered. For each workload, we sample its replica count, inter-arrival time
(which determines the start time), lifetime (which determines the end time) from
the calibrated lifetime distribution, per-pod CPU and memory requests, and
priority level from the distributions described above. This procedure generates
a time-ordered sequence of workload arrivals and removals whose aggregate load
matches the target usage for the corresponding trace configuration.

\paragraph{Trace Replay}

A trace replayer executes the generated traces against the simulated cluster by
translating the trace workloads into Kubernetes API operations. The replayer
first provisions the specified number of virtual nodes and creates
\texttt{PriorityClass} objects for all priority levels. After applying the
initial workloads to reach steady-state load, the replayer issues the newly
arriving trace workloads at their scheduled timestamps. A monitor samples
statistics every~1\,s, recording the number of running and pending pods, pod
deletions, resource usage, and per-pod events (creation and running timestamps).
We exclude initial workloads from the statistics.

For every simulation, we track the following metrics:

\begin{itemize}
  \item \emph{Usage.} At intervals of 1\,s, we record the total requested CPU
  and memory of all running pods, normalised by the total cluster capacity for
  each resource. To capture the bottleneck resource at each time unit, we pick
  the highest between CPU and memory usage and take that metric's average over
  the entire duration of the simulation. Higher values indicate more resource
  usage.

  \item \emph{Scheduling latency.} To quantify scheduling responsiveness, we
  measure the time elapsed between when a pod is submitted to the cluster and
  when it first reaches the running state. We record this latency for each pod
  and report results separately for each priority level. Lower values indicate
  faster scheduling decisions.

  \item \emph{Pod deletions.} To quantify disruption, we track pod deletions
  throughout the simulation. We record each deletion event and report the final
  cumulative count separately for each priority level. Lower values indicate
  less disruptive scheduling behaviour.

  \item \emph{Solver runs.} We record how many times \tool invokes
  optimisations during a simulation. This metric captures the intervention
  frequency of \tool: fewer optimisations imply lower overhead, but may also
  correspond to fewer opportunities for optimisation.

  \item \emph{Plan activations.} We record how many times the optimiser produces
  an improved placement plan that is accepted and enforced by the scheduler.
  This metric captures how often optimisation results in concrete scheduling
  actions, as opposed to solver runs that do not necessarily yield an applicable
  or beneficial plan.
\end{itemize}

\subsection{Experimental Setup}

To stress-test \tool under challenging optimisation conditions, we intentionally
instantiate it with a complete solving approach that tackles the NP-hard pod
placement problem and can therefore require non-negligible computation time
before producing solutions. Concretely, we implement the optimiser by adapting
the classical bin-packing formulation to pod scheduling, encoded and solved
using the CP-SAT constraint programming framework of Google
OR-Tools~\cite{christensen_priority_2025}. Since the scope of this paper is not
the design of the optimisation algorithm itself, but rather the mechanisms that
allow the Kubernetes scheduler to integrate, trigger, and enforce
solver-produced placement plans, we refer the interested reader to the
supplementary material for full details of the optimisation model and its
implementation.

We focus our evaluation on medium-sized clusters, which are (i) representative
of real-world deployments since more than half of companies report operating
clusters of up to 50 nodes
\cite{cncf_cloud_native_computing_foundation_kubernetes_2023-1}, (ii) large
enough to present non-trivial optimisation challenges, and (iii) small enough
that their owners have not conduct extensive ad-hoc optimization
attempts.\footnote{For larger clusters, we expect that a faster but incomplete
technique (e.g., local search) would likely be preferable. Exploring this design
space is left as future work.}

As reported in \cref{tab:trace-params}, we generate traces for two cluster sizes
(16 and 32 nodes), two priority configurations (one priority level and four
priority levels), and three mean inter-arrival time settings to produce a range
of workload patterns, from more bursty (4\,s and 8\,s) to more steady (16\,s)
arrivals, which can impact the optimisation opportunities and the behaviour of
the different \tool modes. We also generate five distinct traces for each
configuration through different random seeds, %
which allows us to capture variability in the results due to randomness in the
workload patterns. We replay every trace under the default Kubernetes scheduler
with and without \tool, evaluating \tool's three triggering modes --
\emph{scheduling-failure}, \emph{periodic}, and \emph{stable-queue} -- across
different timing configurations and for both \emph{blocking} and
\emph{non-blocking} variants.

We run each experiment on an isolated virtual machine with 8 vCPUs and 48 GB of
RAM (Intel Xeon Gold 6130), Ubuntu Linux 24.04, and OR-Tools 9.14.6206.

For reproducibility, the instructions and the code to run the experiments is available at \url{https://zenodo.org/records/19052813} (anonymised for peer review purposes).

\subsection{Results}
\label{sec:results}

\begin{figure}[p]
  \centering
  \includegraphics[height=\textheight - 9\baselineskip]{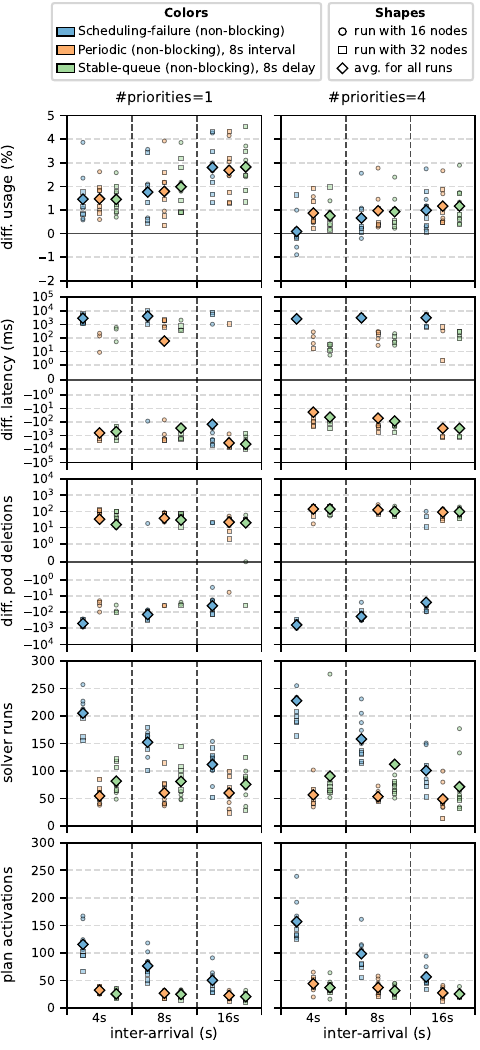}
  \vfill
  \caption{Comparison of \tool modes with \emph{non-blocking}
  variants. Each point shows the difference
  (\tool's mode minus default) for a single run (circles: 16 nodes; squares: 32
  nodes), and diamonds denote the mean difference per trace setting. Left
  column: one priority level; right column: four priority levels. Rows report
  effective usage (\%), scheduling latency (ms), pod deletions, solver runs, and
  plan activations.}
  \label{fig:results-default-preemption=1-blocking=0}
\end{figure}

We start by comparing the \emph{non-blocking} variants of \tool's three
triggering modes: \emph{scheduling-failure}, \emph{periodic} (8\,s interval),
and \emph{stable-queue} (8\,s delay). This comparison is shown in
Figure~\ref{fig:results-default-preemption=1-blocking=0}, which reports paired
differences relative to the default scheduler\footnote{Mathematically, each
measure corresponds to \tool's performance minus the corresponding performance
of the default scheduler.} under two priority configurations: a single priority
level and four priority levels. Each colour denotes an \tool mode, and point
shapes indicate the number of nodes in the cluster.

For each trace setting, we report the mean difference across all runs of that
trace setting as a diamond and each individual run for a specific seed as a
circle (16 nodes) or square (32 nodes). The rows correspond to the following
metrics: (1)~effective usage (higher is better), (2)~scheduling latency in
milliseconds (lower is better), (3)~pod deletions (lower is better), (4)~solver
runs, and (5)~plan activations. The last two rows reports both the runtime
overhead introduced by \tool and how effectively each mode converts optimization
runs into valid plans (the gap between solver runs and plan activations) that
can be activated rather than discarded.

All reported ranges refer to the variation across trace settings, computed from
the mean over the five seeds for each setting. In the following, we first
comment on the results of the different measures (we discuss solver runs and
plan activation together due to their inherent interplay). Then, we briefly
present a set of ablation studies on variables such as (non-)blocking variants, usage of the
DefaultPreemption, parameters of the periodic and stable-queue modes.

\paragraph{Usage}

Almost all \tool modes consistently improve usage
compared to the default scheduler. The improvements are generally more
pronounced for larger inter-arrival times, and when running with a single
priority level. The explanation of these results is straightforward: larger
inter-arrival times provide more opportunities for optimisation. Moreover, with
one priority, the default scheduler struggles to react to suboptimal placements
since it does not reschedule pods after their first placement.

With one priority, \emph{periodic} achieves improvements ranging from
$+1.4\%$ to $+3.0\%$, and \emph{stable-queue} ranges from
$+1.4\%$ to $+2.9\%$, whereas \emph{scheduling-failure} ranges from $+1.2\%$
to $+3.0\%$. With four priority levels, \emph{periodic} yields improvements in
the range $+0.7\%$ to $+1.3\%$ and \emph{stable-queue} ranges from $+0.7\%$ to
$+1.4\%$, while \emph{scheduling-failure} varies from $-0.2\%$ to $+1.2\%$.

Overall, \emph{stable-queue} and \emph{periodic} achieve the best performance
across the settings while \emph{scheduling-failure} is less effective, although
it still improves usage in most settings.

\paragraph{Scheduling latency}

Looking at scheduling latency, \emph{periodic} and
\emph{stable-queue} generally improve upon the default scheduler, whereas
\emph{scheduling-failure} increases latency. We expected this behaviour, since
the latter mode triggers optimisation at each scheduling failure, potentially
causing many solver runs and latency due to solve and plan-enforcement time,
which can also delay other pending pods. Again, improvements are more pronounced
for larger inter-arrival times, and when running with a single priority level.

For a single priority level, \emph{periodic} impacts latency between $-4,235$
and $+83$\,ms (lower is better), and \emph{stable-queue} between $-5,695$ and
$+83$\,ms, whereas \emph{scheduling-failure} between $-1,258$ to $+5,491$\,ms.
With four priority levels, \emph{periodic} ranges from $-439$ to $+65$\,ms and
\emph{stable-queue} ranges from $-351$ to $-3$\,ms, while
\emph{scheduling-failure} varies from $+2,513$ to $+3,329$\,ms. The narrower
improvement range with four priorities is explained by the Kubernetes default
scheduler's \emph{DefaultPreemption} mechanism, which already provides a
priority-aware eviction mechanism leaving less room for improvement.
Specifically, the default scheduler, upon a scheduling failure, calls a
\emph{DefaultPreemption} routine that evicts one or more lower-priority pods to
free resources on a \emph{single} candidate node to accommodate the pending pod.
Looking at the per-priority breakdown (see \cref{tab:latency-defpreempt1-prio4}
in the supplementary material), the latency improvements of \emph{periodic} and
\emph{stable-queue} are driven almost entirely by the lowest-priority class, which sees reductions of up to $1,467$\,ms for \emph{periodic} and up to
$1,216$\,ms for \emph{stable-queue}, while higher-priority classes
experience only marginal changes of tens of milliseconds. In contrast,
\emph{scheduling-failure} increases latency across all priority classes by
similar magnitudes, with the lowest-priority class increases ranging from $+2,465$ to $+3,843$\,ms.

Similarly to the usage metric, \emph{stable-queue} and \emph{periodic} achieve
the best performance across the settings, with a slight edge to
\emph{stable-queue}, which is more consistent across settings, likely derived
from its reactive triggering.

\paragraph{Pod deletions}

Considering pod deletions, as expected from the
results of the previous two metrics, \emph{periodic} and \emph{stable-queue}
most often increase the number of deletions. Their latency gains
derive indeed from pod preemption. Overall, the two modes incur a similar number of
deletions. However, at lower inter-arrival times, \emph{periodic} tends to
trigger fewer deletions than the default scheduler, whereas \emph{stable-queue}
incurs more deletions than the default scheduler across all settings.
Interestingly, \emph{scheduling-failure} reverses the above pattern,
substantially reducing pod deletions w.r.t\@ the default scheduler to gain small
usage improvements at the cost of higher latency.

For a single priority level, \emph{periodic} impacts the number of deletions
(lower is better) between $+2.2$ and $+64.4$, \emph{stable-queue} between $-1.4$
and $+33.2$, whereas \emph{scheduling-failure} between $-528.2$ and $-7.8$. With
four priority levels, \emph{periodic} impacts between $+70.8$ and $+159$
deletions and \emph{stable-queue} between $+73$ and $+145.6$, while
\emph{scheduling-failure} between $-752.6$ and $-19.4$.

\paragraph{Runtime overhead}

Looking at solver runs and plan activations (fourth and fifth rows), \emph{scheduling-failure} triggers far more optimisations than
\emph{periodic} and \emph{stable-queue} -- as expected, given its
scheduling-failure activation behaviour -- while
\emph{stable-queue}
triggers slightly more optimisations than \emph{periodic} -- consistent with the
reactive activation behaviour of the latter mode.

The number of plan activations is generally lower than the number of solver
runs, indicating that not every optimisation yields an accepted plan -- either
because the solver finds no improvement, or because the resulting plan is not
applicable.

\emph{Periodic} stands out as the mode in
which plan activations most closely track solver runs, suggesting that most of
its solver invocations translate into accepted plans.

With a single priority level, \emph{periodic} triggers between $46$ and $76$
solver runs and \emph{stable-queue} triggers between $60$ and $102$ solver runs,
whereas \emph{scheduling-failure} triggers between $110$ and $226$ solver runs.
With four priority levels, \emph{periodic} ranges from $40$ to $59$ solver runs
and \emph{stable-queue} ranges from $50$ to $154$ solver runs, while
\emph{scheduling-failure} varies from $84$ to $266$ solver runs.

\vspace{0.5em}
Overall, the results show that \tool improves cluster usage and, in most
settings, also reduces scheduling latency relative to the default scheduler.
Across the evaluated settings, \emph{periodic} and \emph{stable-queue} are the
strongest modes, achieving the best trade-off between usage and latency
improvements and pod deletions compared to the default scheduler.

\subsubsection{Ablation studies}

We complement the previous experiments and results with another set of
experiments conducted to explore configuration variables that can influence the
behaviour of \tool. We evaluate (i) the difference between \emph{blocking} and
\emph{non-blocking} variants of the modes, (ii) the impact of the
\emph{DefaultPreemption} mechanism, and (iii) the sensitivity of the
\emph{periodic} and \emph{stable-queue} modes to different intervals and
stable-queue idle window durations. For space reasons, we report the detailed
results of these experiments in the supplementary material.

\input{figures/generated/tables/table_blocking_diff_defpreempt=1}

\paragraph{Blocking vs.\ non-blocking}

Table~\ref{tab:blocking-diff-defpreempt1-prioall} shows a comparison between the
\emph{blocking} and \emph{non-blocking} variants. Overall, effective usage is essentially
unchanged ($\le 0.1\%$), and both solver runs and plan activations remain
comparable. The clearest differences are visible for scheduling latency and pod
deletions, where the preferred variant depends on how each trigger mode
interacts with the mean inter-arrival time.

For \emph{scheduling-failure}, the non-blocking variant increases latency at the
two shortest inter-arrival times ($+66$,ms at 4,s, $+303$,ms at 8,s) but reduces
it at 16,s ($-209$,ms). Although it may seem counterintuitive that
\emph{non-blocking} can yield higher latency at shorter inter-arrival times,
this occurs because the cluster state changes frequently during solver
execution, making computed plans more likely to become stale and be discarded.
As a result, the pods that triggered optimisation may remain unschedulable until
a subsequent solver run, increasing their waiting time. In contrast, with longer
inter-arrival times, the plans are more likely to remain valid, allowing
\emph{non-blocking} to proceed with scheduling while the solver runs, which can
reduce latency. At the same time, non-blocking produces substantially fewer pod
deletions at 4\,s ($-142$). This trade-off is consistent with higher trigger
rates and lower inter-arrival times: blocking delays pods until each plan is
available, which helps in keeping plans valid, but the resulting more frequent
rearrangements increase deletions. Non-blocking avoids these deletions by
allowing pods to be scheduled, at the cost of producing stale plans and
increasing latency.

For \emph{periodic}, non-blocking reduces latency at 4\,s ($-133$\,ms) but
increases it at 16\,s ($+245$\,ms), while modestly increasing pod deletions
across all inter-arrival times. The latency improvement for non-blocking under
lower inter-arrival times is expected: proceeding regular scheduling while the
solver runs is beneficial when pods arrive frequently.

For \emph{stable-queue}, the two variants are nearly indistinguishable at the
shorter inter-arrival times, but non-blocking yields a latency improvement at
16\,s ($-266$\,ms). This result matches the design: stable-queue triggers
optimisation only after the scheduling queue remained stable for the configured
delay, so, with longer inter-arrival times, the queue stabilises more
frequently, allowing for more optimisations. In that regard, non-blocking
solving imposes little interference (few arrivals during the solve) while
avoiding the unnecessary scheduling delay introduced by blocking. Pod deletions
remain small and show no consistent trend.

Overall, the results suggest that the non-blocking variant is a reasonable
default choice: it achieves essentially the same effective resource usage as
the blocking variant, while reducing interference with normal scheduling and
exhibiting only small differences in scheduling latency.

\paragraph{Sensitivity to timing parameters in periodic and stable-queue modes}

To assess sensitivity to timing parameters, we vary the triggering
configuration for both periodic and stable-queue modes by testing shorter
(4\,s) and longer (16\,s) values around the 8\,s baseline.
These values are chosen to mirror the mean inter-arrival
times used in our traces (4\,s, 8\,s, and 16\,s), allowing us to study how
optimisation frequency interacts with different workload arrival regimes.
\cref{tab:timing-diff-defpreempt1-blocking0} reports the mean paired
differences between the 8\,s baseline and the alternative configurations.

The dominant effect is on scheduling latency, which
mainly depends on the mean inter-arrival time, while the remaining metrics
largely reflect the expected change in optimisation frequency. For
\emph{periodic}, using a shorter interval (8\,s to 4\,s) yields the largest
latency reduction with 16\,s inter-arrival ($-703$\,ms), whereas using a longer
interval (8\,s to 16\,s) increases latency under lower inter-arrival times
($+187$\,ms at 4\,s). For \emph{stable-queue}, latency shifts are smaller: a
shorter idle window (8\,s to 4\,s) slightly reduces latency at 4--8\,s, while
a longer idle window (8\,s to 16\,s) increases latency across inter-arrival
times.

Therefore, as expected, shorter intervals/windows increase the number of solver
invocations, which can reduce latency, but this higher optimisation frequency
typically increases pod deletions because more frequent plan activations lead to
more evictions and relocations.

\input{figures/generated/tables/table_timing_diff_defpreempt=1_blocking=0}
\input{figures/generated/tables/table_defpreempt_diff_blocking=0}

\paragraph{Impact of \emph{DefaultPreemption}}

To isolate the impact of \tool's preemption and plan-enforcement
mechanisms from Kubernetes' built-in behaviour, we also repeat the
experiments with \emph{DefaultPreemption} disabled in the default scheduler.
This configuration prevents Kubernetes from evicting pods opportunistically
upon a scheduling failure, ensuring that any observed evictions and
reallocations originate solely from \tool.

As shown in Table~\ref{tab:defpreempt-diff-blocking0}, having
\emph{DefaultPreemption} enabled primarily shifts the trade-off between latency and pod
deletions, with only small effects on usage and number of solver runs. Across
all three trigger modes and inter-arrival times, effective usage changes by $\le
0.1\%$. In contrast, scheduling latency consistently decreases when
\emph{DefaultPreemption} is enabled: the latency difference (enabled $-$
disabled) is negative in every setting, ranging from $-459$\,ms to $-2,194$\,ms
for \emph{scheduling-failure}, $-728$\,ms to $-1,558$\,ms for \emph{periodic},
and $-684$\,ms to $-2,185$\,ms for \emph{stable-queue}, with the largest
reductions occurring at longer inter-arrival times. This latency improvement is
explained by the built-in DefaultPreemption heuristic that
allows the scheduler to quickly evict lower-priority pods to make room for
pending pods, rather than waiting for a solver-produced plan to be computed and
enforced. However, this improvement comes at the cost of more pod deletions
because \emph{DefaultPreemption} evicts pods greedily on a single node without
considering the global cluster state. As a result, enabling
\emph{DefaultPreemption} increases deletions in all settings (from $+84.8$ up to
$+533.9$), with the largest increases typically observed at the shorter
inter-arrival times. Solver runs and plan activations change little in
comparison.

Furthermore, with multiple priorities, the latency gains are concentrated in the
lowest-priority:
nearly all improvements come from the
lowest-priority pods, while the three higher-priority classes change only
marginally, consistent with \emph{DefaultPreemption} already protecting
high-priority pods.

%% file: figures/generated/tables/table_blocking_diff_defpreempt=1.tex
\begin{table*}[htbp]
\centering
\caption{Mean difference (non-blocking $-$ blocking) per inter-arrival time, aggregated over node counts and all priority configurations, with DefaultPreemption enabled.}
\label{tab:blocking-diff-defpreempt1-prioall}
\adjustbox{max width=\textwidth}{%
\begin{tabular}{l w{c}{2.5em} w{c}{2.5em} w{c}{2.5em} w{c}{2.5em} w{c}{2.5em} w{c}{2.5em} w{c}{2.5em} w{c}{2.5em} w{c}{2.5em}}
\toprule
\multirow{2}{*}{\textbf{Metric}} & \multicolumn{3}{c}{\makecell{\textbf{Scheduling-failure}\\(non-blocking\,$-$\,blocking)}} & \multicolumn{3}{c}{\makecell{\textbf{Periodic, 8s interval}\\(non-blocking\,$-$\,blocking)}} & \multicolumn{3}{c}{\makecell{\textbf{Stable-queue, 8s delay}\\(non-blocking\,$-$\,blocking)}} \\
\cmidrule(lr){2-4}\cmidrule(lr){5-7}\cmidrule(lr){8-10}
 & \llap{Inter-arrival =\,}4s & 8s & 16s & 4s & 8s & 16s & 4s & 8s & 16s \\
\midrule
Diff. usage (\%) & +0.1 & +0.0 & +0.0 & +0.0 & +0.0 & -0.1 & +0.0 & +0.0 & +0.0 \\
Diff. latency (ms) & +66 & +303 & -209 & -133 & +32 & +245 & +30 & -3 & -266 \\
Diff. pod deletions & -142.0 & -10.2 & -4.2 & +20.6 & +9.2 & +2.6 & +11.4 & -2.0 & +11.4 \\
Diff. solver runs & +4 & +1 & +1 & +1 & +0 & +2 & +6 & +5 & +0 \\
Diff. plan activations & -7 & +2 & +1 & +0 & +2 & +0 & -1 & +0 & +1 \\
\bottomrule
\end{tabular}%
}
\end{table*}

%% file: figures/generated/tables/table_timing_diff_defpreempt=1_blocking=0.tex
\begin{table*}[htbp]
\centering
\caption{Mean difference between 8s baseline and other timing variants per inter-arrival time, aggregated over node counts and priority configurations, for non-blocking runs with DefaultPreemption enabled.}
\label{tab:timing-diff-defpreempt1-blocking0}
\adjustbox{max width=\textwidth}{%
\begin{tabular}{l w{c}{2.5em} w{c}{2.5em} w{c}{2.5em} w{c}{2.5em} w{c}{2.5em} w{c}{2.5em} w{c}{2.5em} w{c}{2.5em} w{c}{2.5em} w{c}{2.5em} w{c}{2.5em} w{c}{2.5em}}
\toprule
\multirow{3}{*}{\textbf{Metric}} & \multicolumn{6}{c}{\textbf{Periodic}} & \multicolumn{6}{c}{\textbf{Stable-queue}} \\
\cmidrule(lr){2-7}\cmidrule(lr){8-13}
 & \multicolumn{3}{c}{Interval: 8s $\to$ 4s} & \multicolumn{3}{c}{Interval: 8s $\to$ 16s} & \multicolumn{3}{c}{Delay: 8s $\to$ 4s} & \multicolumn{3}{c}{Delay: 8s $\to$ 16s} \\
\cmidrule(lr){2-4}\cmidrule(lr){5-7}\cmidrule(lr){8-10}\cmidrule(lr){11-13}
 & \llap{Inter-arrival =\,}4s & 8s & 16s & 4s & 8s & 16s & 4s & 8s & 16s & 4s & 8s & 16s \\
\midrule
Diff. usage (\%) & +0.1 & +0.2 & +0.1 & -0.2 & -0.1 & +0.0 & +0.1 & +0.1 & +0.3 & -0.2 & -4.6 & +0.0 \\
Diff. latency (ms) & +21 & -217 & -703 & +187 & +22 & -88 & -57 & -22 & +29 & +105 & +78 & +76 \\
Diff. pod deletions & +43.4 & +24.9 & +17.4 & -45.8 & -34.1 & -16.7 & +37.2 & +23.8 & +1.9 & -45.0 & -118.2 & -17.9 \\
Diff. solver runs & +38 & +32 & +21 & -24 & -24 & -22 & +82 & +32 & +39 & -47 & -47 & -31 \\
Diff. plan activations & +21 & +13 & +8 & -15 & -11 & -6 & +16 & +7 & +5 & -12 & -9 & -4 \\
\bottomrule
\end{tabular}%
}
\end{table*}

%% file: figures/generated/tables/table_defpreempt_diff_blocking=0.tex
\begin{table*}[htbp]
\centering
\caption{Mean difference (DefaultPreemption enabled $-$ disabled) per inter-arrival time, aggregated over node counts and all priority configurations, for non-blocking runs.}
\label{tab:defpreempt-diff-blocking0}
\adjustbox{max width=\textwidth}{%
\begin{tabular}{l w{c}{2.5em} w{c}{2.5em} w{c}{2.5em} w{c}{2.5em} w{c}{2.5em} w{c}{2.5em} w{c}{2.5em} w{c}{2.5em} w{c}{2.5em}}
\toprule
\multirow{2}{*}{\textbf{Metric}} & \multicolumn{3}{c}{\makecell{\textbf{Scheduling-failure}\\(enabled\,$-$\,disabled)}} & \multicolumn{3}{c}{\makecell{\textbf{Periodic, 8s interval}\\(enabled\,$-$\,disabled)}} & \multicolumn{3}{c}{\makecell{\textbf{Stable-queue, 8s delay}\\(enabled\,$-$\,disabled)}} \\
\cmidrule(lr){2-4}\cmidrule(lr){5-7}\cmidrule(lr){8-10}
 & \llap{Inter-arrival =\,}4s & 8s & 16s & 4s & 8s & 16s & 4s & 8s & 16s \\
\midrule
Diff. usage (\%) & -0.1 & +0.0 & +0.0 & +0.0 & -0.1 & -0.1 & -0.1 & +0.0 & +0.1 \\
Diff. latency (ms) & -459 & -1501 & -2194 & -728 & -1033 & -1558 & -684 & -1291 & -2185 \\
Diff. pod deletions & +260.4 & +158.8 & +84.8 & +525.7 & +297.9 & +144.5 & +533.9 & +289.9 & +149.4 \\
Diff. solver runs & +0 & +3 & +0 & +3 & +2 & +4 & +12 & +17 & -8 \\
Diff. plan activations & +0 & +5 & +0 & +1 & +2 & +1 & +0 & +1 & +1 \\
\bottomrule
\end{tabular}%
}
\end{table*}

%% file: sections/05_related_work.tex
\section{Related Work}
\label{sec:related_work}

We structure the comparison of our proposal with related work following the
categorisation from recent surveys~\cite{senjab_survey_2023,
ahmad_container_2022, rejiba_custom_2023, carrion_kubernetes_2023} along three
main directions: \emph{(meta)heuristics}, \emph{machine learning} (ML), and
\emph{solver-based} approaches.

\paragraph{(Meta)Heuristics}

Several works use (meta)heuristics for pod placement, either at scheduling time
or by rescheduling pods to consolidate resources or reduce load
imbalance~\cite{rejiba_custom_2023}.

Moritz de Carvalho Neto et al.~\cite{moritz_de_carvalho_neto_dynamic_2025}
propose Kubernetes Scheduling Extension (KSE), which introduces periodic,
metric-driven load balancing by migrating pods from over- to under-loaded nodes,
based on CPU and memory usage.
Zhang et al.~\cite{zhang_atasl_2025} introduce ATASL, a custom Kubernetes
scheduler that augments scheduling with application-type awareness.
Fu et al.~\cite{fu_procon_2019} propose ProCon, a scheduler that adds scheduling
logic to both the control-plane and the worker nodes. In ProCon, workers monitor
per-container progress, and the control plane estimates remaining runtimes to
rank nodes and reduce resource contention.
Wei-Guo et al.~\cite{wei-guo_research_2018} propose a cost-aware scheduling
model using metaheuristics that combines particle swarm optimisation and ant
colony optimisation.

Overall, (meta)heuristic-based approaches can find good placements quickly but
offer no optimality guarantees and require careful parameter
tuning~\cite{qawqzeh_review_2021}. Integration-wise, we did not find a proposal
able to co-exist with the default schedular as \tool does. 
However, since \tool is solver agnostic, it can exploit these kind of (meta)heuristics
assuming they can be incorporated into the external solver to compute the placement plans.
We leave this integration as future work.

\paragraph{Machine Learning}

Machine learning approaches for Kubernetes scheduling fall into two
subcategories: \emph{deep learning forecasting}, which predicts indicators
(e.g., upcoming usage levels, queue length, estimated pod runtimes) and feeds
them into scoring or scheduling constraints; \emph{deep reinforcement learning},
which trains an agent to learn a scheduling policy that maximises rewards such
as usage~\cite{carrion_kubernetes_2023}.

Yang et al.~\cite{yang_design_2019} exemplify the deep learning forecasting
approach by combining a grey model with long short-term memory to predict
resource usage dynamics. Their design augments the cluster with monitoring and
prediction modules to generate forecasts, which are then consumed by a
scheduling module that periodically reallocates resources based on the predicted
trends.
Jian et al.~\cite{jian_drs_2024} instead exemplify reinforcement learning with DRS,
which is a Deep Q-Network-based (DQN) scheduler that comes as an alternative
Kubernetes scheduler.

Similarly to (meta)heuristics, we did not find ML-based methods that integrate into Kubernetes
as \tool does. ML-based approaches also often face practical limitations due to the need for substantial training
data and careful model tuning. Moreover, their performance can degrade under
workload shifts. Clearly, better forecasts and
learned policies could be used as a guidance for producing better placements plans, and \tool could integrate them as well by using them as part of the external solver.

\paragraph{Solver-based Approaches}

There are a few proposals that incorporate exact solvers for cluster-wide
placement decisions we were inspired by.

Boreas~\cite{hacid_boreas_2021} is a scheduler that targets
optimal placements subject to hard constraints by batching newly arriving pods
and solving a placement optimisation problem for each batch. Boreas is probably the closes work to ours, but differently than \tool they execute the optimisation outside the default scheduler running the Zephyrus2 optimiser~\cite{abraham_zephyrus2_2016} at every decision point.

SAGE~~\cite{luca_sage_2023} operates at pre-deployment
stage: given an application description and a catalogue of cloud offers it computes both a cost-minimising node-pool selection and a
placement plan but lacks real-time knowledge of resources consumed by Kubernetes at
deployment time.

DCM~\cite{suresh_dcm_2020} is an alterantive scheduler
(architecture) where one can express cluster-management policies as SQL
constraints stored in a relational database. A compiler translates the SQL
schema and constraint views into an encoder that builds a optimisation model,
solved with Google OrTools' CP-SAT backend. At runtime, DCM's scheduler invokes
this encoder,
constructs the
optimisation model, and returns actions applied via the API server.

The considered solver-based proposals either replace the default scheduler or
indirectly operate it. A
common denominator of all these approaches is the renowned trade-off of
solver-based scheduling: computing online cluster-wide placements requires
solving an NP-hard combinatorial problem, which can endure high latency. \tool
builds on a solver-based approach, but it is an upstream-compatible plugin that
smoothly integration with Kubernetes default scheduler to minimize the number of calls to the external solver and use as much as possible the default scheduler's logic to make placement decisions.

%% file: sections/06_conclusions.tex
\section{Conclusions}
\label{sec:conclusions}

We presented \tool, a Kubernetes scheduler plugin that enables an external
optimisation solver to drive cluster-wide placement decisions while reusing the
default Kubernetes scheduler as-is. Unlike prior work that replaces or sidesteps
the default scheduler, \tool extends it via the Kubernetes Scheduling Framework,
achieving clean integration with no modifications to upstream scheduler code.

Our experiments show that \tool consistently improves resource usage without any major drawback
and the periodic and stable-queue modes
reduce scheduling latency in most settings. 
The non-blocking
variant achieves resource usage comparable to the blocking variant while
reducing interference with normal scheduling, which makes it a sound default
choice. Our ablation studies confirm that these results are robust across a
range of trigger intervals and idle-window durations.

Our evaluation instantiated \tool with a complete, NP-hard bin-packing
formulation solved via Google OR-Tools CP-SAT, deliberately stressing the
integration mechanism rather than optimising for speed. A natural and direct
extension is to replace this solver with faster, incomplete approaches -- such
as local search heuristics or metaheuristic methods -- which would reduce
per-invocation latency and expand the operating coverage of \tool to larger
clusters and higher-arrival-rate regimes. In this regard, we view metaheuristic
solvers as promising drop-in replacements one can integrate without changes to
the plugin architecture.

The current design is solver-agnostic but does not exploit runtime signals
beyond the cluster snapshot. Integrating machine-learning forecasts of workload
arrival rates, pod lifetimes, or resource demands as soft constraints or
objective modifiers could allow the solver to make anticipatory placement
decisions, reducing reactive disruption and further improving long-term resource
efficiency.

%% file: appendix/algorithms.tex
\section{Algorithms}
\label{appendix:algorithms}

In this appendix, we detailed the algorithms that are used in our system design.
\input{appendix/algorithms/optimisation_flow}
\newpage
\input{appendix/algorithms/hook_procedures}
\newpage
\input{appendix/algorithms/background_loop}
\newpage
\input{appendix/algorithms/convergence_monitor}

%% file: appendix/algorithms/optimisation_flow.tex
\subsection{Optimisation Flow Procedure}
\label{appendix:optimisation-flow}

Algorithm~\ref{alg:optimization-flow} presents the entry point of the optimisation flow. The procedure is atomically guarded so at most one run is in progress at a time~(line~2). In \emph{blocking} mode, scheduling is paused before the snapshot~(lines~4--5); in \emph{non-blocking} mode, it is paused only after validation~(lines~12--13). The snapshot is taken~(line~6) and the run is skipped if no pods are pending~(lines~7--8). The optimiser runs under a wall-clock timeout~(line~9) and the resulting plan is validated against a fresh snapshot~(lines~10--11). If valid, the plan is installed, making it visible to scheduling workers~(line~14), evictions are carried out~(lines~15--16), planned pods are activated~(line~17), and an asynchronous convergence monitor is started~(line~18). \textsc{TeardownRun}~(lines~20--22) clears the plan, unblocks scheduling, and requeues held-back pods.

\begin{algorithm}[H]
  \caption{Optimisation Flow Procedure.}
  \label{alg:optimization-flow}
    \begin{algorithmic}[1]
        \Procedure{OptimizationFlow}{\textit{blockingOptimizer}, \textit{timeoutOptimizer}}
            \If{\(\Call{OptimizationInProgress} \)}
                \State \Return \Comment{already optimising; skip this run}
            \EndIf
            \If{\(\textit{blockingOptimizer}\)}
                \State \(\Call{BlockScheduling}{ }\) \Comment{blocking: block scheduling before taking snapshot}
            \EndIf
            \State \(\textit{state} \gets \Call{StateSnapshot}{ }\)
            \If{\(\textbf{not}\ \Call{PodsPending}{\textit{state}}\)}
                \State \Return \(\Call{TeardownRun}{ }\) \Comment{no pending pods}
            \EndIf
            \State \(\textit{plan} = (placements,moves,evictions) \gets \Call{RunOptimizer}{\textit{state}, \textit{timeoutOptimizer}}\)
            \If{\(\textbf{not}\ \Call{ValidPlan}{\textit{plan}, \textit{state}}\)}
                \State \Return \(\Call{TeardownRun}{ }\) \Comment{plan failed validation}
            \EndIf
            \If{\(\textbf{not}\ \textit{blockingOptimizer}\)}
                \State \(\Call{BlockScheduling}{ }\) \Comment{non-blocking: first block scheduling here before enforcing plan}
            \EndIf
            \State \(\Call{InstallPlan}{\textit{plan}}\) \Comment{make plan visible to workers}
            \If{\(\textbf{not}\ \Call{EvictPods}{\textit{evictions} \cup \textit{moves}}\)}

                \State \Return \(\Call{TeardownRun}{ }\) \Comment{evict failed or timeout}
            \EndIf
            \State \(\Call{ActivatePlannedPods}{placements}\) \Comment{enqueue pending pods selected for placement or moved}
            \State \(\Call{StartAsync}{\Call{ConvergenceMonitor}{ }}\) \Comment{start async.\ monitor plan convergence in background}
            \State \Return \(\textit{plan}\)
        \EndProcedure

        \par\vspace{6pt}\nointerlineskip
        \Procedure{TeardownRun}{ }
            \State \(\Call{UninstallPlan}{ }\)
            \State \(\Call{UnblockScheduling}{ }\)
        \EndProcedure
    \end{algorithmic}
\end{algorithm}

%% file: appendix/algorithms/hook_procedures.tex
\subsection{Hook Procedures for Plan Enforcement}
\label{appendix:hook-procedures}

Algorithms~\ref{alg:hooks-preenqueue}--\ref{alg:hooks-reserve-unreserve} present the hook procedures used to enforce a plan through the scheduling workers.

\paragraph{Admission control (PreEnqueue).} While a plan is active, \textsc{PreEnqueue} blocks pods not covered by the plan~(line~2) and records them for later requeuing~(lines~3--4).

\begin{algorithm}[H]
    \caption{PreEnqueue Hook: Admission Control.}
    \label{alg:hooks-preenqueue}
    \begin{algorithmic}[1]
        \Procedure{PreEnqueue}{\textit{pod}, \textit{plan}}
            \If {\(\Call{IsPlanActive}{ }\ \textbf{and}\ \textbf{not}\ \Call{IsPodPlanned}{\textit{pod}, \textit{plan}}\)}
                \State \(\Call{AddBlockedPod}{\textit{pod}}\)
                \State \(\Call{BlockPod}{\textit{pod}}\)
            \EndIf
        \EndProcedure
    \end{algorithmic}
\end{algorithm}

\paragraph{Node constraining (PreFilter).} \textsc{PreFilter} restricts candidate nodes to those permitted by the plan for non-protected pods~(lines~2--3); otherwise regular scheduling proceeds.

\begin{algorithm}[H]
    \caption{PreFilter Hook: Node Constraining.}
    \label{alg:hooks-prefilter}
    \begin{algorithmic}[1]
        \Procedure{PreFilter}{\textit{pod}, \textit{nodes}, \textit{plan}}
            \If{\(\Call{IsPlanActive}{ }\ \textbf{and}\ \textbf{not}\ \Call{IsPodProtected}{\textit{pod}}\)}
                \State \(\Call{ConstrainToNodes}{\textit{pod}, nodes, \textit{plan}}\)
            \EndIf
        \EndProcedure
    \end{algorithmic}
\end{algorithm}

\paragraph{Concurrency-safe reservations (Reserve/Unreserve).} \textsc{Reserve} enforces per-workload quotas for controller-managed pods~(line~4); quota exhaustion or a pod-node mismatch causes rejection~(line~5). \textsc{Unreserve} restores consumed quota if the scheduling attempt later fails~(line~7).

\begin{algorithm}[H]
    \caption{Reserve and Unreserve Hooks: Concurrency-Safe Scheduling.}
    \label{alg:hooks-reserve-unreserve}
    \begin{algorithmic}[1]
        \Procedure{Reserve}{\textit{pod}, \textit{node}, \textit{plan}}
            \If{\(\textbf{not}\ \Call{IsPlanActive}{ }\ \textbf{or}\ \Call{IsPodProtected}{\textit{pod}}\) \par
                \hspace{\algorithmicindent}\hspace{\algorithmicindent}\(\textbf{or}\ \textbf{not}\ \Call{HasWorkloadOwner}{\textit{pod}}\)}
                \State \Return \Comment{no reservation}
            \EndIf
            \If{\(\textbf{not}\ \Call{MakeReservation}{\textit{pod}, \textit{node}}\)}
                \State \Return \(\Call{RejectPod}{\textit{pod}}\) \Comment{over quota or pod-node mismatch}  
            \EndIf
        \EndProcedure
        \Procedure{Unreserve}{\textit{pod}, \textit{node}, \textit{plan}}
            \State \(\Call{ReturnReservation}{\textit{pod}, \textit{node}, \textit{plan}}\)
        \EndProcedure
    \end{algorithmic}
\end{algorithm}

%% file: appendix/algorithms/background_loop.tex
\subsection{Trigger Modes Procedures}
\label{sec:background-loop}

This subsection details the procedures for the three trigger modes supported by \tool: \emph{scheduling-failure}, which invokes optimisation synchronously in the scheduling path; \emph{periodic}, which runs it at fixed intervals; and \emph{stable-queue}, which defers it until the pending queue stabilises.

\subsubsection{Scheduling-Failure Mode}

Algorithm~\ref{alg:hooks-postfilter} presents the \emph{scheduling-failure} trigger. When a pod fails to find a feasible node, the \emph{PostFilter} hook invokes the optimisation flow synchronously~(line~2) and, if a plan is produced, nominates a node for the triggering pod~(line~3).

\begin{algorithm}[H]
    \caption{PostFilter Hook Procedure for Scheduling-Failure Trigger Mode.}
    \label{alg:hooks-postfilter}
    \begin{algorithmic}[1]
        \Procedure{PostFilter}{\textit{pod}}
            \State \(\textit{plan} \gets \Call{OptimizationFlow}{ }\)
            \State \(\Call{NominateNode}{\textit{plan}}\)
        \EndProcedure
    \end{algorithmic}
\end{algorithm}

\subsubsection{Periodic and Stable-Queue Modes}

Algorithm~\ref{alg:bgLoop} presents the loop shared by the \emph{periodic} and \emph{stable-queue} modes. The loop sleeps for the configured interval~(line~3), snapshots the cluster state~(line~4), and skips the tick if an optimisation is already in progress~(line~5). In \emph{stable-queue} mode, an in-progress run is torn down when the pending set changes~(lines~6--7). The tick is also skipped if the state has not changed since the last solved run~(lines~9--10). In \emph{stable-queue} mode, the pending set must additionally have remained stable for the configured stability window~(lines~11--12). If all guards pass, the optimisation flow is started asynchronously~(line~13).

\begin{algorithm}[H]
\caption{Background Loop for Periodic and Stable-Queue Trigger Modes.}
\label{alg:bgLoop}
\begin{algorithmic}[1]
  \Procedure{BackgroundLoop}{\textit{mode}, \textit{timeoutInterval}, \textit{timeoutStability}}
    \Loop
      \State \(\Call{Sleep}{\textit{timeoutInterval}}\)
      \State $\textit{state} \gets \Call{StateSnapshot}{ }$
      \If{\(\Call{IsOptimizationInProgress}{ }\)}
        \If{$\textit{mode}=\textsc{StableQueue}$ \par
            \hspace{\algorithmicindent}\hspace{\algorithmicindent}$\textbf{and}\ \Call{PendingChanged}{\textit{state}}$}
          \State \(\Call{TeardownRun}{ }\)
        \EndIf
        \State \(\textbf{continue}\) \Comment{plan being enforced or pending set changed}
      \EndIf
      \If{\(\Call{AlreadySolved}{\textit{state}}\)}
        \State \(\textbf{continue}\) \Comment{state unchanged since last solve}
      \EndIf

      \If{$\textit{mode}=\textsc{StableQueue}$ \par
          \hspace{\algorithmicindent}\hspace{\algorithmicindent}$\textbf{and not}\ \Call{QueueStable}{\textit{timeoutStability}}$}
        \State \(\textbf{continue}\) \Comment{wait for stability window}
      \EndIf

      \State \(\Call{StartAsync}{\Call{OptimizationFlow}{ }}\)
    \EndLoop
  \EndProcedure
\end{algorithmic}
\end{algorithm}

%% file: appendix/algorithms/convergence_monitor.tex
\subsection{Convergence Procedure for Plan Monitoring}
\label{appendix:convergence-monitor}

Algorithm~\ref{alg:planCompletionWatcher1a} presents the convergence monitor, an asynchronous routine started after a plan becomes active. The monitor records its start time~(line~2) and periodically sleeps for the configured check interval~(line~4). If the plan deadline is exceeded, the run is torn down~(lines~5--6). Otherwise, the cluster state is snapshotted~(line~7) and convergence is evaluated via two conditions~(line~8): \textsc{StablePodsPlacedOk} checks identity-stable pods by verifying that each non-terminated pod is observed on its designated node; missing or terminated pods are treated as satisfied~(lines~12--16). \textsc{UnstablePodsQuotasOk} checks identity-unstable (controller-managed) pods at the workload level, verifying that per-node quotas have been fully consumed or no active replicas remain~(lines~19--24). If both conditions hold, the run is torn down and regular scheduling resumes~(line~9).

\begin{algorithm}[H]
    \caption{Convergence Procedure for Plan Monitoring.}
    \label{alg:planCompletionWatcher1a}
    \begin{algorithmic}[1]
        \Procedure{ConvergenceMonitor}{\textit{plan}, \textit{cvrgInterval}, \textit{timeoutPlan}}
            \State \(\textit{startTime} \gets \Call{Now}{ }\)
            \Loop
                \State \(\Call{Sleep}{\textit{cvrgInterval}}\)

                \If{\((\Call{Now}{ } - \textit{startTime}) \ge \textit{timeoutPlan}\)}
                    \State \Return \(\Call{TeardownRun}{ }\) \Comment{plan timed out}
                \EndIf

                \State \textit{state} $\gets$ \Call{StateSnapshot}{ }

                \If{\(\Call{StablePodsPlacedOk}{\textit{state}, \textit{plan}}\) \par
                    \hspace{\algorithmicindent}\(\textbf{and}\ \Call{UnstablePodsQuotasOk}{\textit{state}, \textit{plan}}\)}
                    \State \Return \(\Call{TeardownRun}{ }\) \Comment{plan converged}
                \EndIf
            \EndLoop
        \EndProcedure

        \par\vspace{6pt}\nointerlineskip
        \Procedure{StablePodsPlacedOk}{\textit{state}, \textit{placeByName}}
            \For{\((\textit{pod}, \textit{wantNode})\) \textbf{in} \textit{placeByName}}
                \If{\(\Call{IsPodGone}{\textit{pod}}\)}
                    \State \(\textbf{continue}\) \Comment{satisfied; terminated or deleted}
                \EndIf
                \If{\(\textbf{not}\ \Call{IsPodOnNode}{\textit{pod}, \textit{wantNode}}\)}
                    \State \Return \textbf{false} \Comment{not satisfied; pod not on desired node}
                \EndIf
            \EndFor
            \State \Return \textbf{true} \Comment{all pods placed on desired nodes}
        \EndProcedure

        \par\vspace{6pt}\nointerlineskip
        \Procedure{UnstablePodsQuotasOk}{\textit{state}, \textit{wkQuotas}}
            \For{\((\textit{wk}, \textit{remaining})\) \textbf{in} \textit{wkQuotas}}
                \If{\(\textit{remaining} \le 0\ \textbf{or}\ \textbf{not}\ \Call{HasWkActivePods}{\textit{wk}}\)}
                    \State \(\textbf{continue}\) \Comment{workload quota satisfied or workload scaled down or deleted}
                \EndIf
                \If{\(\Call{HasPendingPods}{\textit{wk}}\)}
                    \State \Return \textbf{false} \Comment{not satisfied; pending pods with workload quota remaining}
                \EndIf
            \EndFor
            \State \Return \textbf{true} \Comment{all workload quotas satisfied}
        \EndProcedure
    \end{algorithmic}
\end{algorithm}

%% file: appendix/public_trace_analysis.tex
\section{Analysis and Plots of Public Traces}
\label{appendix:public-trace-analysis}

The two analyzed public traces--the Alibaba GPU
trace~\cite{alibaba_alibabacluster-trace-gpu-v2025_2025} and Google Cluster Data
2019~\cite{google_developers_googleclusterdata2019_2025}--show similar
heavy-tailed patterns in inter-arrival times, workload lifetimes, and resource
requests. This is consistent with prior
analyses~\citep{reiss_heterogeneity_2012}. Resource requests are also skewed,
although the Alibaba request histograms appear more discretized. Despite these
differences, we model inter-arrivals, lifetimes, and CPU/memory requests using
the same bounded Pareto distribution to keep the generator consistent and
simple.

For discrete-valued attributes, we use truncated geometric distributions.
Priorities--available in the Google trace--are well approximated by a geometric
distribution, capturing that most workloads concentrate in a small number of low
priority levels while higher levels occur with diminishing frequency. Replica
counts are not available in either public trace, so we likewise model them
geometrically, reflecting the common pattern that most workloads run with few
replicas while a smaller fraction scale out to many.

\begin{figure}[H]
    \centering
    \includegraphics[width=1\linewidth, clip, trim=0.2cm 0.3cm 0.2cm 0.2cm]{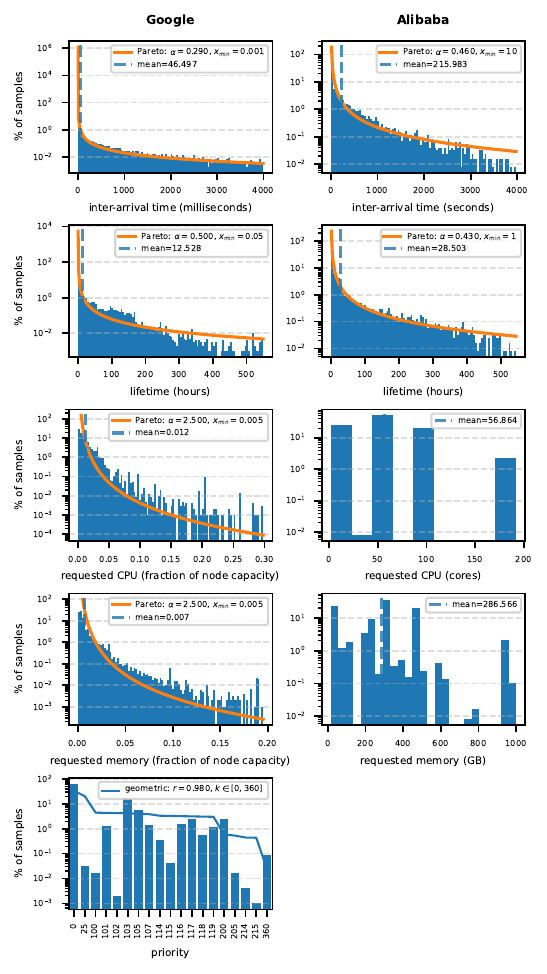}
    \caption{Histograms of pods' inter-arrival times, lifetimes, CPU/memory
    requests, and priorities from publicly available cluster traces. The orange
    lines show fitted Pareto distributions for inter-arrival times, lifetimes,
    and resource requests. Priorities are fitted with a geometric distribution.
    No priorities are available in the Alibaba trace.}
    \label{fig:public-trace-histograms}
\end{figure}

Figure~\ref{fig:public-trace-histograms} shows histograms from the public traces
along with the fitted distributions used to inform the synthetic generator.

%% file: appendix/optimisation_and_solver.tex
\section{Optimisation Model and External Solver}

In this appendix we detail the optimisation model we designed and the external solver we use to implement it.

\subsection{Optimisation problem}

We solve the optimisation problem incrementally, in a loop that iterates over
all priorities. At every priority tier, we run a solver twice to 1) maximise the
number of placed pods, and 2) minimise the number of allocated pods we need to
move.

Notation: we use $n \in N$ to denote nodes. We define with $n.\text{ram}$ and
$n.\text{cpu}$ the capacities of $n$. For each pod $p \in P$, let $p.\text{ram}$
and $p.\text{cpu}$ be its resource requests, $p.\text{priority} \in [0,
pr_{\max}]$ its priority level (with lower values denoting higher priority), and
$p.\text{where}$ the index of the node where the pod is currently placed
(\(p.\text{where}=0\) if the pod is not scheduled). The binary variables $x_{i,
j} \in \{0, 1\}$ indicate whether pod $p_i$ is assigned to node $n_j$.

\algrenewcommand{\algorithmiccomment}[1]{%
  \hfill \textit{\color{gray}\(\triangleright\) #1}}

\begin{algorithm}[t]
\caption{\label{alg:opt}Optimisation Algorithm.}
\begin{algorithmic}[1]
\State $m \gets \text{Model}()$
\For{$pr \in \text{range}(p_{\max}+1)$}
    \State $m.\text{add\_constraint}(\text{bin\_packing\_constraints}(pr))$
    \State\Comment{(1) maximise number of placed pods}
    \State $\text{metric} \gets \sum_{p_i \in P: p_i.\text{priority} \leq pr} \sum_{n_j \in N} x_{i,j}$
    \State $\text{sol} \gets m.\text{max}(\text{metric}, \text{get\_timeout}())$
    \If{$\text{sol}.\text{status} = \text{OPTIMAL}$}
        \State $m.\text{add\_constraint}(\text{metric} = \text{sol}(\text{metric}))$
    \Else
        \State $m.\text{add\_constraint}(\text{metric} \geq \text{sol}(\text{metric}))$
    \EndIf
    \State\Comment{(2) minimise number of pod moves}
    \State $\text{metric} \gets$
    \State \phantom{A}$\sum_{\substack{p_i \in P: p_i.\text{priority} \leq pr \\ \wedge\ p_{i}.\text{ where} \neq 0}} \left( \sum_{n_j \in N} x_{i,j} + 2 x_{i, p_i.\text{where}} \right)$
    \State $\text{sol} \gets m.\text{max}(\text{metric}, \text{get\_timeout}())$
    \If{$\text{sol}.\text{status} = \text{OPTIMAL}$}
        \State $m.\text{add\_constraint}(\text{metric} = \text{sol}(\text{metric}))$
    \Else
        \State $m.\text{add\_constraint}(\text{metric} \leq \text{sol}(\text{metric}))$
    \EndIf
\EndFor
\State \Return $\text{sol}$
\end{algorithmic}
\end{algorithm}

We report in \cref{alg:opt} the pseudocode of the optimisation loop, where a
model $m$ collects constraints and supports a solver that maximises a given
metric.

At Line 1, we initialise the model without constraints. Then, for each priority
level \(pr\) from the highest (0) to the lowest (\(p_{max}\)) we add
(multi-dimensional) bin-packing constraints (Line 3) to restrict solutions to
valid placements, and use these to (1) maximise the pods allocated up to
priority \(pr\) (Lines 5--10), and (2) to minimise the removal of deployed pods
(Lines 12--18).

Constraints
impose that the sum of the
requested resources by the pod with priority up to \(pr\) are within the
respective node's RAM and CPU capacities, 
and ensures assignment of a pod with priority up to \(pr\) to at most one
node.

At Line 6 of \cref{alg:opt}, we maximise the number of pods within the current
priority by maximising of the metric
\[\sum_{p_i \in P:\, p_i.\mathrm{priority}\le pr}\sum_{n_j \in N} x_{i,j}.\]

If the solver finds an \texttt{OPTIMAL} solution, we update the model by adding
a constraint that imposes to find future solutions with exactly the same amount
of pods up to priority \(pr\) (Line 8). If the solver did not terminate within
the allocated time, thus not proving that the solution is optimal but providing
only a \texttt{FEASIBLE} solution, we add a constraint that imposes to find
solutions with at least the same amount of pods up to priority \(pr\) (Line 10).

In the second stage, we minimise the disruptions caused by avoidable pod
evictions. To do so, we minimise the evictions of allocated pods up to priority
\(pr\) by asking the solver, at Line 14, to maximise the following metric:
\[
\sum_{p_i \in P:\, p_i.\mathrm{priority}\le pr \land p_i.\mathrm{where}\ne 0} \left(\sum_{n_j \in N} x_{i,j} + 2x_{i,p_i.\mathrm{where}}\right)
\]
where we give a weight of 2 to the pods that remain in place (i.e., assigned to
the same node they are already allocated onto) and a weight of 1 to the pods
to move to another node.

As the optimisation problem is NP-hard, we chose to use a fixed wall-clock
timeout $T_{\mathrm{total}}$, which limits the total solving time across all
priority tiers. To avoid that a single priority tier exhausts the entire time
budget, thus compromising the possibility to find a solution involving also
lower priorities pods, we dedicate a fraction $\alpha \in [0,1]$ of the total
time to each priority tier, leaving the remaining $(1-\alpha)
T_{\mathrm{total}}$ time for the opportunistic termination of the solver
executions. Moreover, for every priority, we split in half the reserved time
between the two solving phases (maximisation of pods and minimisation of
evictions). In particular, in \cref{alg:opt}, we use function
$\texttt{get\_timeout}$ at every solver invocation (Line 6 and 14) to compute
the available time budget. Formally, assuming that \texttt{unused} is the amount
of unreserved time yet to consume, $\alpha \in [0,1]$
\[\texttt{get\_timeout}() = \alpha T_{\mathrm{total}}/({p_{max}}+1) +
\texttt{unused} \]

\subsection{Solver}

We implement our approach by using OR-Tools~\cite{PF24} -- an open-source suite
developed by Google for optimisation tasks using, among other approaches, a
lazy-constraint approach. Notably, OR-Tools is the winner of the latest MiniZinc
Challenge~\cite{minizinc_challenge}, a competition evaluating the performance of
constraint programming solvers across diverse benchmarks. In particular, we use
the OR-Tools' CP-SAT Python API, which runs several complementary search
strategies in parallel. Since, unlike popular SMT solvers (e.g., Z3), CP-SAT
does not support incremental push/pop of constraints, we re-solve the model
after each place/move step. To try to speed up the search, we use CP-SAT hints
to warm-start the next solve by providing the current assignment.

We integrate the solver and its optimisation model into \tool via the Scheduling Framework, where \tool calls a Python script that invokes OR-Tools deployed inside the scheduler container.

%% file: appendix/results.tex
\section{Results}
\label{appendix:results}

This appendix complements the results in Section~\ref{sec:results} with: (i)
figures for \emph{blocking} variants and with \emph{DefaultPreemption} disabled,
(ii) sensitivity figures for the \emph{periodic} interval and
\emph{stable-queue} idle window, (iii) tables reporting the mean and standard
deviation of each metric for every mode and trace setting, and (iv) a comparison
of \tool's modes with \emph{DefaultPreemption} enabled vs. disabled.

The appendix is structured as follows:
Section~\ref{appendix:results-default-preemption=1} presents results with
\emph{DefaultPreemption} enabled.
Section~\ref{appendix:results-default-preemption=0} presents results with
\emph{DefaultPreemption} disabled. Finally,
Section~\ref{appendix:results-default-preemption-enabled-vs-disabled} compares
\tool's modes with \emph{DefaultPreemption} enabled vs. disabled.

\subsection{Results with DefaultPreemption enabled}
\label{appendix:results-default-preemption=1}

This subsection presents additional results for \tool's modes when running with
\emph{DefaultPreemption} enabled.

\begin{figure*}[!t]
  \centering
  \includegraphics[width=0.58\linewidth]{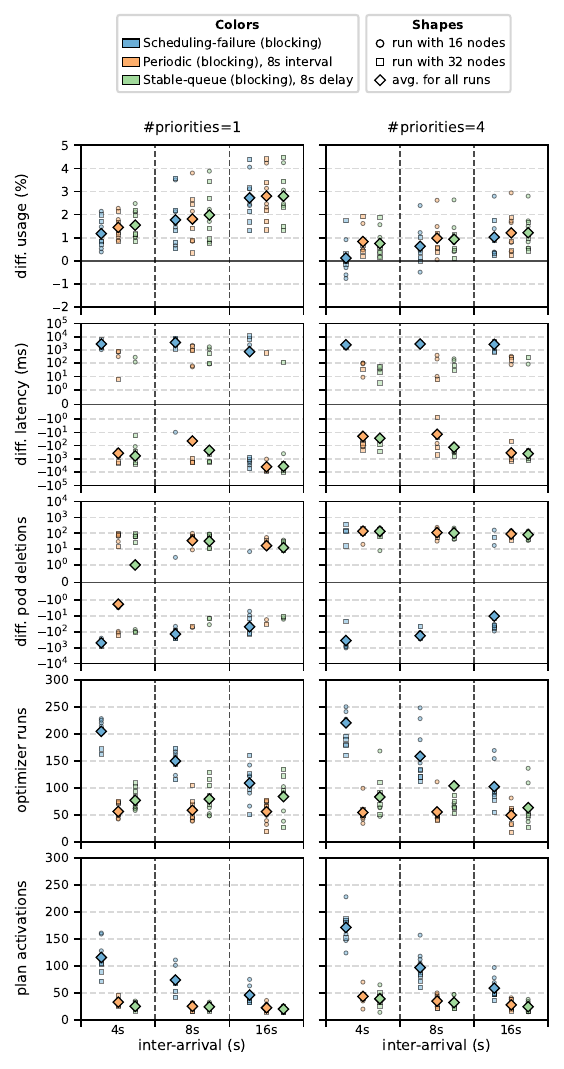}
  \caption{Comparison of \tool modalities with \emph{blocking} variants and
  \emph{DefaultPreemption} enabled. Each point shows the difference (\tool's
  mode minus default) for a single run (circles: 16 nodes; squares: 32 nodes),
  and diamonds denote the mean difference per trace setting. Left column: one
  priority level; right column: four priority levels. Rows report effective
  usage (\%), scheduling latency (ms), pod deletions, solver runs, and plan
  activations.}
  \label{fig:results-default-preemption=1-blocking=1}
\end{figure*}

Figure~\ref{fig:results-default-preemption=1-blocking=1} shows the results for
\emph{blocking} variants of \tool's modes with \emph{DefaultPreemption} enabled.

\begin{figure*}[!t]
  \centering
  \includegraphics[width=0.58\linewidth]{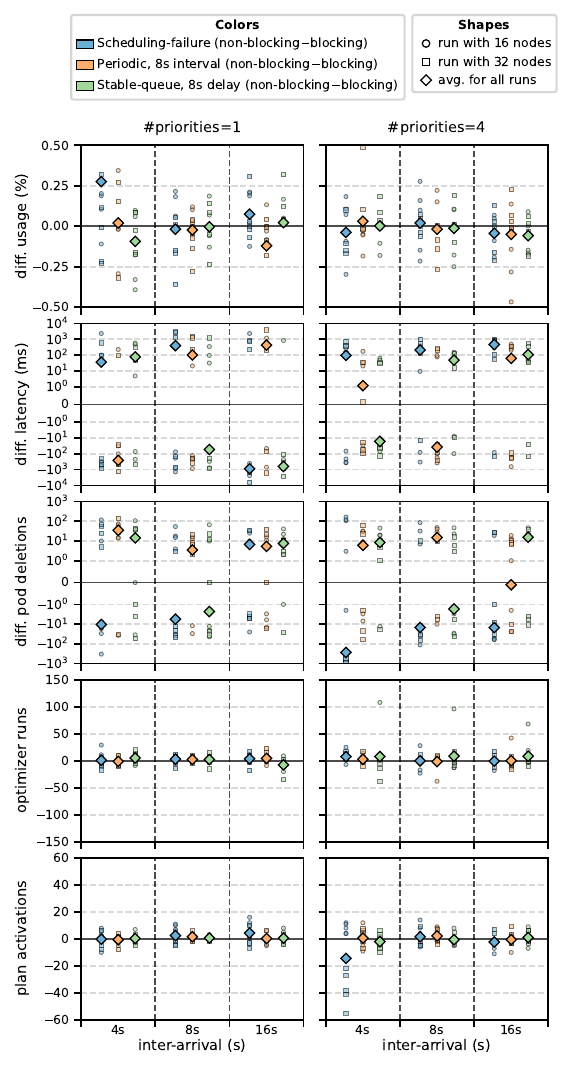}
  \caption{Comparison of \tool modalities (\emph{non-blocking} vs.
  \emph{blocking}; \emph{DefaultPreemption} enabled). Each point shows the
  difference (\emph{non-blocking} minus \emph{blocking}) for a single run
  (circles: 16 nodes; squares: 32 nodes), and diamonds denote the mean
  difference per trace setting (mean across five seeds). Left column: one
  priority level; right column: four priority levels. Rows report effective
  usage (\%), scheduling latency (ms), pod deletions, solver runs, and plan
  activations.}
  \label{fig:blocking-diffs-default-preemption=1}
\end{figure*}

Figure~\ref{fig:blocking-diffs-default-preemption=1} compares \tool's modes with
\emph{non-blocking} vs. \emph{blocking} variants when running with
\emph{DefaultPreemption} enabled.

\begin{figure*}[!t]
  \centering
  \includegraphics[width=0.58\linewidth]{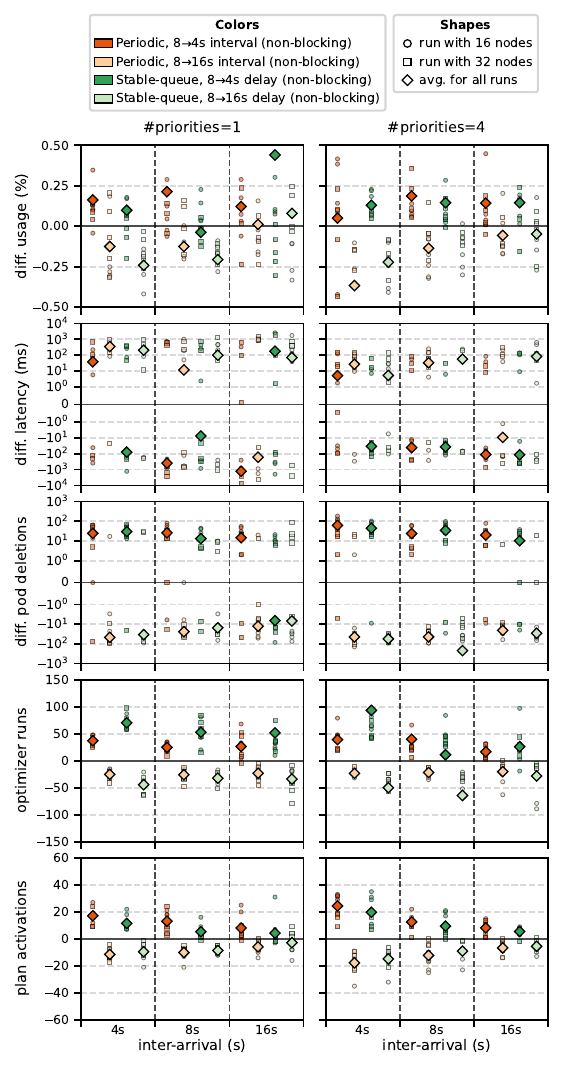}
  \caption{Comparison of \tool modalities (\emph{non-blocking};
  \emph{DefaultPreemption} enabled) with a sensitivity analysis of the
  \emph{periodic} interval and \emph{stable-queue} idle window. The sensitivity
  setting compares a shorter (4\,s) and longer (16\,s) interval/idle window
  against the default 8\,s configuration (8\,s$\rightarrow$4\,s and
  8\,s$\rightarrow$16\,s). Each point shows the difference (\tool minus default)
  for a single run (circles: 16 nodes; squares: 32 nodes), and diamonds denote
  the mean difference per trace setting (mean across five seeds). Left column:
  one priority level; right column: four priority levels. Rows report effective
  usage (\%), scheduling latency (ms), pod deletions, solver runs, and plan
  activations.}
  \label{fig:periodic-stable-queue-sensitivity-default-preemption=1-blocking=0}
\end{figure*}

\begin{figure*}[!t]
  \centering
  \includegraphics[width=0.58\linewidth]{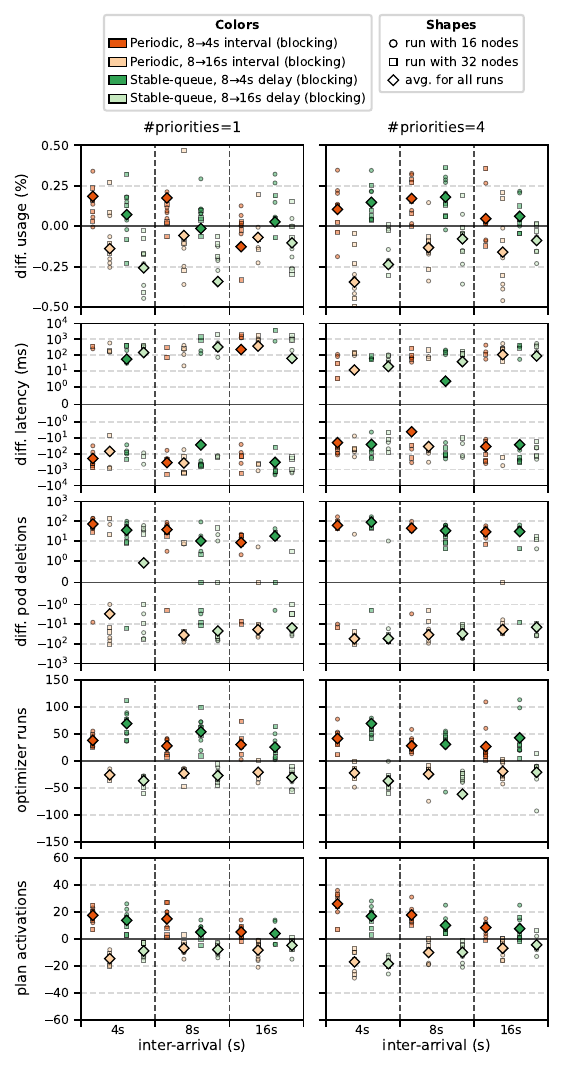}
  \caption{Comparison of \tool modalities (\emph{blocking};
  \emph{DefaultPreemption} enabled) with a sensitivity analysis of the
  \emph{periodic} interval and \emph{stable-queue} idle window. The sensitivity
  setting compares a shorter (4\,s) and longer (16\,s) interval/idle window
  against the default 8\,s configuration (8\,s$\rightarrow$4\,s and
  8\,s$\rightarrow$16\,s). Each point shows the difference (\tool minus default)
  for a single run (circles: 16 nodes; squares: 32 nodes), and diamonds denote
  the mean difference per trace setting (mean across five seeds). Left column:
  one priority level; right column: four priority levels. Rows report effective
  usage (\%), scheduling latency (ms), pod deletions, solver runs, and plan
  activations.}
  \label{fig:periodic-stable-queue-sensitivity-default-preemption=1-blocking=1}
\end{figure*}

Figures~\ref{fig:periodic-stable-queue-sensitivity-default-preemption=1-blocking=0} and
\ref{fig:periodic-stable-queue-sensitivity-default-preemption=1-blocking=1}
shows how the \emph{periodic} interval and \emph{stable-queue} idle window
affect usage, scheduling latency, solver runs and plan activations. Each mode is
evaluated with shorter (4\,s) and longer (16\,s) settings relative to the
default 8\,s, using the \emph{non-blocking} and \emph{blocking} variants,
respectively, with \emph{DefaultPreemption} enabled.

Tables~\ref{tab:usage-defpreempt1-prio1}--\ref{tab:plan_activations-defpreempt1-prio4},
report the mean and standard deviation of each metric for each mode and trace
setting when running with \emph{DefaultPreemption} enabled containing all runs
(\emph{non-blocking} and \emph{blocking} variants) for both priority
configurations (one and four priority levels).

\input{figures/generated/tables/table_defpreempt=1_priorities=1_usage}
\input{figures/generated/tables/table_defpreempt=1_priorities=4_usage}
\input{figures/generated/tables/table_defpreempt=1_priorities=1_latency}
\input{figures/generated/tables/table_defpreempt=1_priorities=4_latency}
\input{figures/generated/tables/table_defpreempt=1_priorities=1_deletions}
\input{figures/generated/tables/table_defpreempt=1_priorities=4_deletions}
\input{figures/generated/tables/table_defpreempt=1_priorities=1_optimizer_runs}
\input{figures/generated/tables/table_defpreempt=1_priorities=4_optimizer_runs}
\input{figures/generated/tables/table_defpreempt=1_priorities=1_plan_activations}
\input{figures/generated/tables/table_defpreempt=1_priorities=4_plan_activations}

\subsection{Results with DefaultPreemption disabled}
\label{appendix:results-default-preemption=0}

This subsection reports results for \tool's modes with \emph{DefaultPreemption}
disabled. The same modes are run, as in Section~\ref{sec:results}, but the
default scheduler is prevented from preempting pods in \emph{PostFilter}. This
isolates the impact of \tool's preemption mechanism.

\begin{figure*}[!t]
  \centering
  \includegraphics[width=0.58\linewidth]{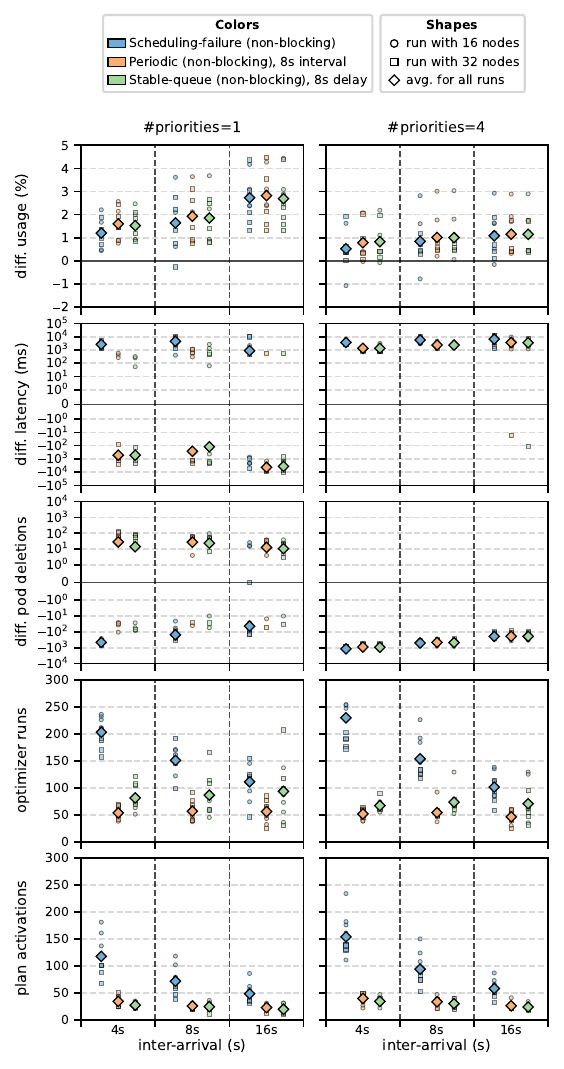}
  \caption{Comparison of \tool modalities with \emph{non-blocking} variants and
  \emph{DefaultPreemption} disabled. Each point shows the difference (\tool's
  mode minus default) for a single run (circles: 16 nodes; squares: 32 nodes),
  and diamonds denote the mean difference per trace setting. Left column: one
  priority level; right column: four priority levels. Rows report effective
  usage (\%), scheduling latency (ms), pod deletions, solver runs, and plan
  activations.}
  \label{fig:results-default-preemption=0_blocking=0}
\end{figure*}

\begin{figure*}[!t]
  \centering
  \includegraphics[width=0.58\linewidth]{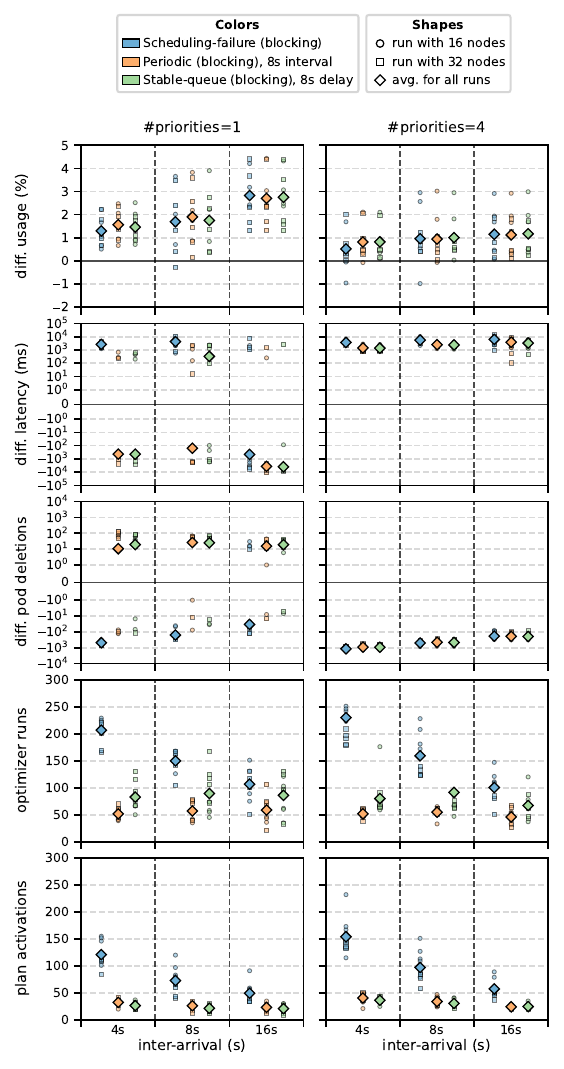}
  \caption{Comparison of \tool modalities with \emph{blocking} variants and
  \emph{DefaultPreemption} disabled. Each point shows the difference (\tool's
  mode minus default) for a single run (circles: 16 nodes; squares: 32 nodes),
  and diamonds denote the mean difference per trace setting. Left column: one
  priority level; right column: four priority levels. Rows report effective
  usage (\%), scheduling latency (ms), pod deletions, solver runs, and plan
  activations.}
  \label{fig:results-default-preemption=0_blocking=1}
\end{figure*}

Figures~\ref{fig:results-default-preemption=0_blocking=0} and
\ref{fig:results-default-preemption=0_blocking=1} summarises results when
running \tool's modes in \emph{non-blocking} and \emph{blocking} variants with
\emph{DefaultPreemption} disabled, respectively.

\begin{figure*}[!t]
  \centering
  \includegraphics[width=0.58\linewidth]{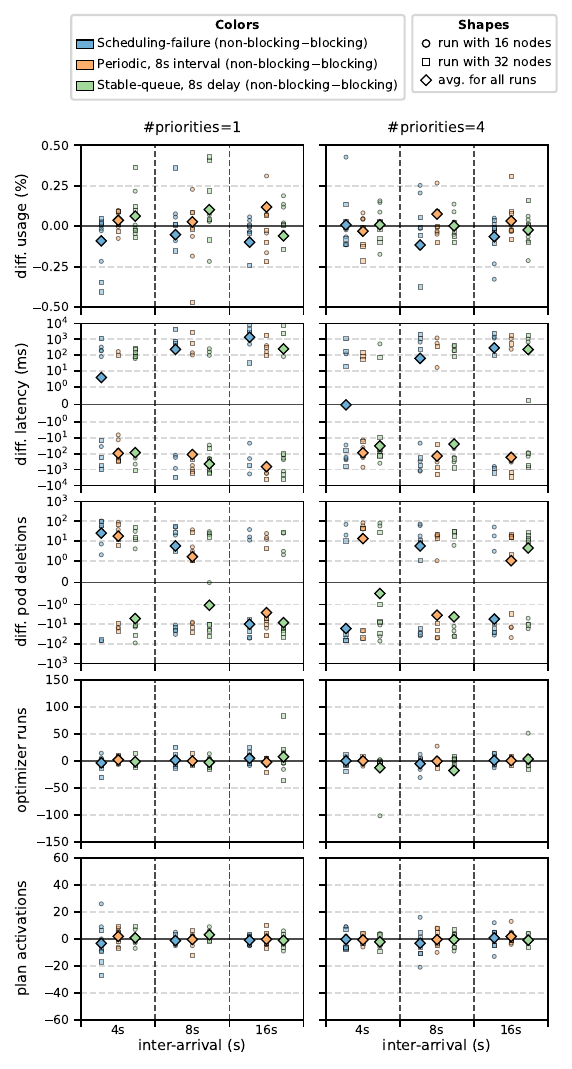}
  \caption{Comparison of \tool modalities (\emph{non-blocking} vs.
  \emph{blocking}; \emph{DefaultPreemption} disabled). Each point shows the
  difference (\emph{non-blocking} minus \emph{blocking}) for a single run
  (circles: 16 nodes; squares: 32 nodes), and diamonds denote the mean
  difference per trace setting (mean across five seeds). Left column: one
  priority level; right column: four priority levels. Rows report effective
  usage (\%), scheduling latency (ms), pod deletions, solver runs, and plan
  activations.}
  \label{fig:blocking-diffs-default-preemption=0}
\end{figure*}

Figure~\ref{fig:blocking-diffs-default-preemption=0} compares \tool's modes with
\emph{non-blocking} vs. \emph{blocking} variants when running with
\emph{DefaultPreemption} disabled.

\begin{figure*}[!t]
  \centering
  \includegraphics[width=0.58\linewidth]{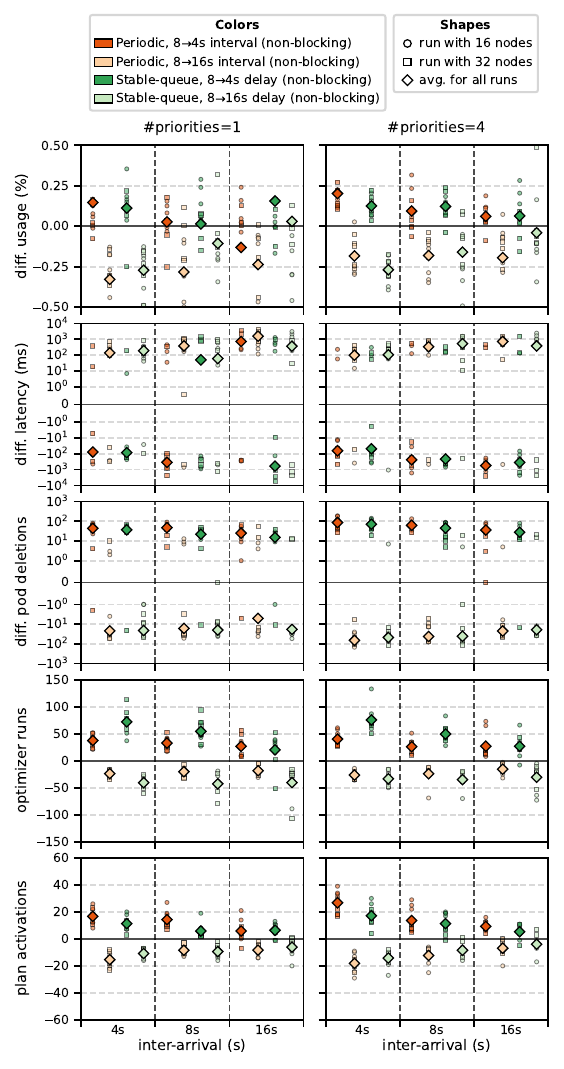}
  \caption{Comparison of \tool modalities (\emph{non-blocking};
  \emph{DefaultPreemption} disabled) with a sensitivity analysis of the
  \emph{periodic} interval and \emph{stable-queue} idle window. The sensitivity
  setting compares a shorter (4\,s) and longer (16\,s) interval/idle window
  against the default 8\,s configuration (8\,s$\rightarrow$4\,s and
  8\,s$\rightarrow$16\,s). Each point shows the difference (\tool minus default)
  for a single run (circles: 16 nodes; squares: 32 nodes), and diamonds denote
  the mean difference per trace setting (mean across five seeds). Left column:
  one priority level; right column: four priority levels. Rows report effective
  usage (\%), scheduling latency (ms), pod deletions, solver runs, and plan
  activations.}
  \label{fig:periodic-stable-queue-sensitivity-default-preemption=0-blocking=0}
\end{figure*}

\begin{figure*}[!t]
  \centering
  \includegraphics[width=0.58\linewidth]{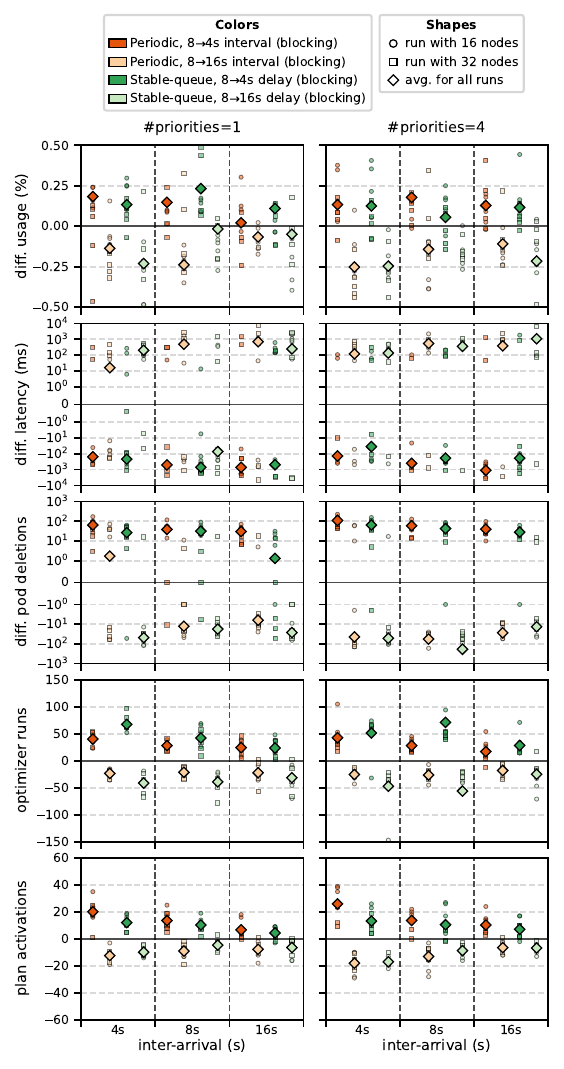}
  \caption{Comparison of \tool modalities (\emph{blocking};
  \emph{DefaultPreemption} disabled) with a sensitivity analysis of the
  \emph{periodic} interval and \emph{stable-queue} idle window. The sensitivity
  setting compares a shorter (4\,s) and longer (16\,s) interval/idle window
  against the default 8\,s configuration (8\,s$\rightarrow$4\,s and
  8\,s$\rightarrow$16\,s). Each point shows the difference (\tool minus default)
  for a single run (circles: 16 nodes; squares: 32 nodes), and diamonds denote
  the mean difference per trace setting (mean across five seeds). Left column:
  one priority level; right column: four priority levels. Rows report effective
  usage (\%), scheduling latency (ms), pod deletions, solver runs, and plan
  activations.}
  \label{fig:periodic-stable-queue-sensitivity-default-preemption=0-blocking=1}
\end{figure*}

Figures~\ref{fig:periodic-stable-queue-sensitivity-default-preemption=0-blocking=0} and
\ref{fig:periodic-stable-queue-sensitivity-default-preemption=0-blocking=1}
reports the sensitivity of the \emph{periodic} and \emph{stable-queue} modes to
their interval and idle window, respectively, when running with
\emph{non-blocking} and \emph{blocking} variants, respectively.

Tables~\ref{tab:usage-defpreempt0-prio1}--\ref{tab:plan_activations-defpreempt0-prio4},
report the mean and standard deviation of each metric for each mode and trace
setting when running with \emph{DefaultPreemption} disabled containing all runs
(\emph{non-blocking} and \emph{blocking} variants) for both priority
configurations (one and four priority levels).

\input{figures/generated/tables/table_defpreempt=0_priorities=1_usage}
\input{figures/generated/tables/table_defpreempt=0_priorities=4_usage}
\input{figures/generated/tables/table_defpreempt=0_priorities=1_latency}
\input{figures/generated/tables/table_defpreempt=0_priorities=4_latency}
\input{figures/generated/tables/table_defpreempt=0_priorities=1_deletions}
\input{figures/generated/tables/table_defpreempt=0_priorities=4_deletions}
\input{figures/generated/tables/table_defpreempt=0_priorities=1_optimizer_runs}
\input{figures/generated/tables/table_defpreempt=0_priorities=4_optimizer_runs}
\input{figures/generated/tables/table_defpreempt=0_priorities=1_plan_activations}
\input{figures/generated/tables/table_defpreempt=0_priorities=4_plan_activations}

\subsection{Results for DefaultPreemption Enabled vs.\ Disabled}
\label{appendix:results-default-preemption-enabled-vs-disabled}

\begin{figure*}[!t]
  \centering
  \includegraphics[width=0.58\linewidth]{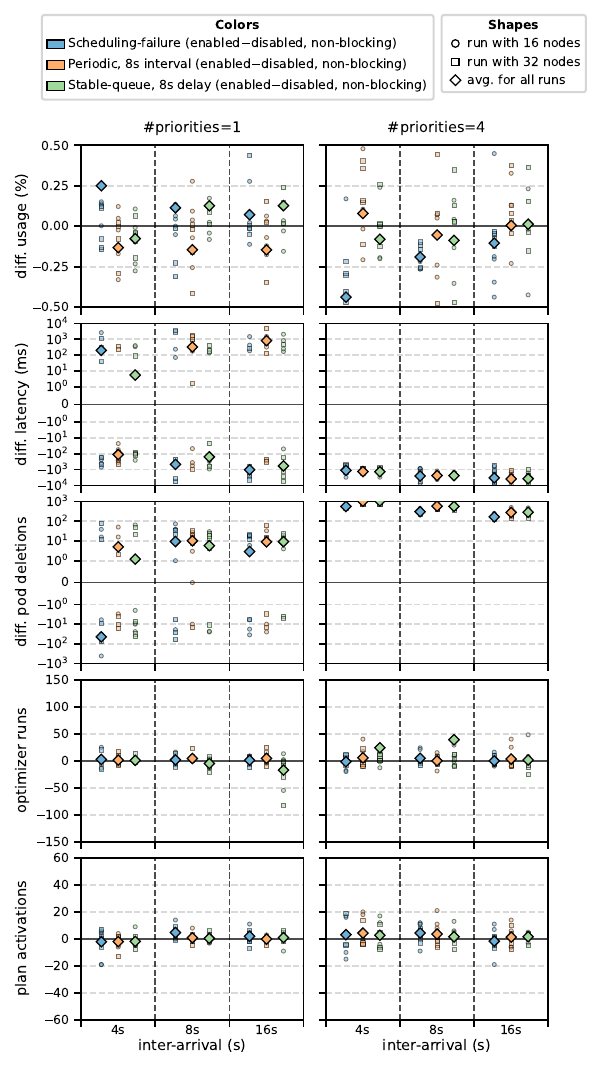}
  \caption{Comparison of \tool modalities (\emph{DefaultPreemption} enabled vs. disabled) with \emph{non-blocking} variants. Each point shows the difference (\emph{DefaultPreemption} enabled minus \emph{DefaultPreemption} disabled) for a single run (circles: 16 nodes; squares: 32 nodes), and diamonds denote the mean difference per trace setting (mean across five seeds). Left column: one priority level; right column: four priority levels. Rows report effective usage (\%), scheduling latency (ms), pod deletions, solver runs, and plan activations.}
  \label{fig:defpreempt-diffs-blocking=0}
\end{figure*}

Figure~\ref{fig:defpreempt-diffs-blocking=0} compares \tool's modes with
\emph{DefaultPreemption} enabled vs. disabled when running with
\emph{non-blocking} and \emph{blocking} variants, respectively.

\begin{figure*}[!t]
  \centering
  \includegraphics[width=0.58\linewidth]{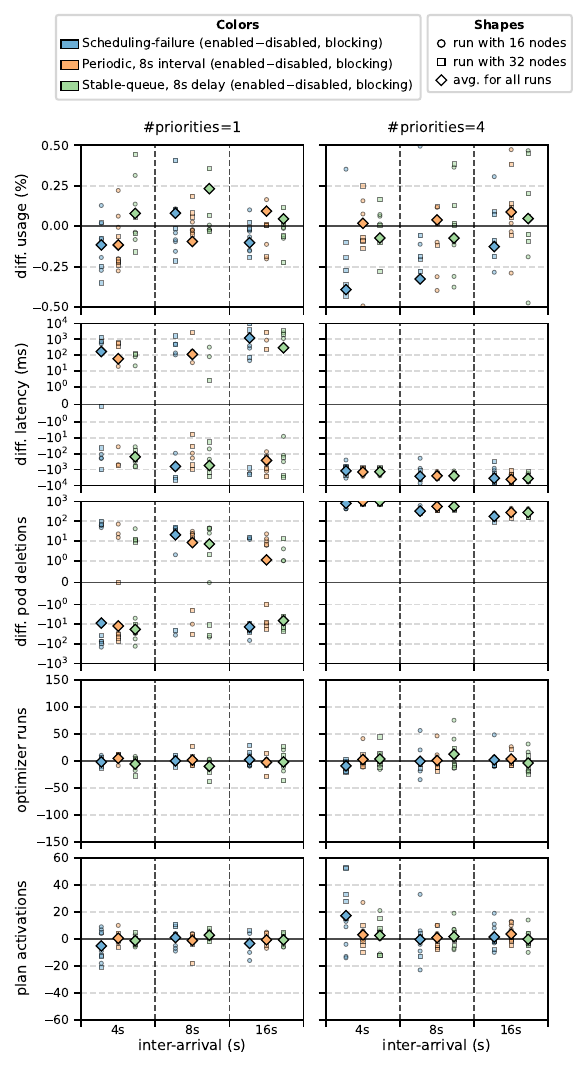}
  \caption{Comparison of \tool modalities (\emph{DefaultPreemption} enabled vs. disabled) with \emph{blocking} variants. Each point shows the difference (\emph{DefaultPreemption} enabled minus \emph{DefaultPreemption} disabled) for a single run (circles: 16 nodes; squares: 32 nodes), and diamonds denote the mean difference per trace setting (mean across five seeds). Left column: one priority level; right column: four priority levels. Rows report effective usage (\%), scheduling latency (ms), pod deletions, solver runs, and plan activations.}
  \label{fig:defpreempt-diffs-blocking=1}
\end{figure*}

%% file: figures/generated/tables/table_defpreempt=1_priorities=1_usage.tex
\begin{table*}[htbp]
\centering
\caption{Mean paired differences in effective resource usage (\%) between the plugin with DefaultPreemption enabled and the default scheduler for runs with one priority (mean ± std).}
\label{tab:usage-defpreempt1-prio1}
\adjustbox{max width=\textwidth, max totalheight=0.9\textheight}{%
\begin{tabular}{l w{c}{5.0em} w{c}{5.0em} w{c}{5.0em} w{c}{5.0em} w{c}{5.0em} w{c}{5.0em}}
\toprule
\multirow{2}{*}{\textbf{Trigger Mode}} & \multicolumn{3}{c}{\# Nodes =\,16} & \multicolumn{3}{c}{\# Nodes =\,32} \\
\cmidrule(lr){2-4}\cmidrule(lr){5-7}
 & \llap{Inter-arrival =\,}4s & 8s & 16s & 4s & 8s & 16s \\
\midrule
Scheduling-failure (blocking) & $+1.1\pm0.7$ & $+1.9\pm1.1$ & $+2.9\pm0.7$ & $+1.2\pm0.6$ & $+1.7\pm1.2$ & $+2.5\pm1.2$ \\
Scheduling-failure (non-blocking) & $+1.7\pm1.4$ & $+2.0\pm1.0$ & $+3.0\pm0.7$ & $+1.2\pm0.4$ & $+1.6\pm1.2$ & $+2.6\pm1.3$ \\
Periodic (blocking), 4s interval & $+1.7\pm0.8$ & $+2.2\pm1.1$ & $+3.0\pm0.8$ & $+1.5\pm0.5$ & $+1.8\pm1.3$ & $+2.3\pm1.0$ \\
Periodic (non-blocking), 4s interval & $+1.6\pm0.7$ & $+2.2\pm1.1$ & $+3.1\pm0.8$ & $+1.6\pm0.6$ & $+1.8\pm1.3$ & $+2.5\pm1.3$ \\
Periodic (blocking), 8s interval & $+1.5\pm0.6$ & $+2.1\pm1.1$ & $+3.0\pm0.8$ & $+1.4\pm0.5$ & $+1.5\pm1.0$ & $+2.6\pm1.3$ \\
Periodic (non-blocking), 8s interval & $+1.5\pm0.8$ & $+2.1\pm1.1$ & $+3.0\pm0.7$ & $+1.4\pm0.4$ & $+1.5\pm0.9$ & $+2.4\pm1.3$ \\
\bottomrule
\end{tabular}%
}
\end{table*}

%% file: figures/generated/tables/table_defpreempt=1_priorities=4_usage.tex
\begin{table*}[htbp]
\centering
\caption{Mean paired differences in effective resource usage (\%) between the plugin with DefaultPreemption enabled and the default scheduler for runs with 4 priorities (mean ± std). p1 is the lowest priority.}
\label{tab:usage-defpreempt1-prio4}
\adjustbox{max width=\textwidth, max totalheight=0.9\textheight}{%
\begin{tabular}{l w{c}{5.0em} w{c}{5.0em} w{c}{5.0em} w{c}{5.0em} w{c}{5.0em} w{c}{5.0em}}
\toprule
\multirow{2}{*}{\textbf{Trigger Mode}} & \multicolumn{3}{c}{\# Nodes =\,16} & \multicolumn{3}{c}{\# Nodes =\,32} \\
\cmidrule(lr){2-4}\cmidrule(lr){5-7}
 & \llap{Inter-arrival =\,}4s & 8s & 16s & 4s & 8s & 16s \\
\midrule
Scheduling-failure (blocking) & $-0.2\pm0.7$ & $+0.8\pm1.1$ & $+1.2\pm1.0$ & $+0.4\pm0.8$ & $+0.4\pm0.4$ & $+0.8\pm0.6$ \\
Scheduling-failure (non-blocking) & $-0.2\pm0.7$ & $+0.9\pm1.1$ & $+1.2\pm1.0$ & $+0.3\pm0.7$ & $+0.4\pm0.4$ & $+0.8\pm0.7$ \\
Periodic (blocking), 4s interval & $+1.0\pm0.6$ & $+1.4\pm1.0$ & $+1.5\pm0.9$ & $+0.9\pm0.7$ & $+0.9\pm0.4$ & $+1.0\pm0.5$ \\
Periodic (non-blocking), 4s interval & $+0.9\pm0.7$ & $+1.5\pm1.1$ & $+1.6\pm1.0$ & $+0.9\pm0.7$ & $+0.8\pm0.3$ & $+1.0\pm0.6$ \\
Periodic (blocking), 8s interval & $+0.9\pm0.5$ & $+1.2\pm0.9$ & $+1.5\pm1.0$ & $+0.8\pm0.7$ & $+0.8\pm0.3$ & $+0.9\pm0.7$ \\
Periodic (non-blocking), 8s interval & $+0.8\pm0.5$ & $+1.3\pm0.9$ & $+1.3\pm0.9$ & $+0.9\pm0.7$ & $+0.7\pm0.4$ & $+1.0\pm0.6$ \\
\bottomrule
\end{tabular}%
}
\end{table*}

%% file: figures/generated/tables/table_defpreempt=1_priorities=1_latency.tex
\begin{table*}[htbp]
\centering
\caption{Mean paired differences in scheduling latency (ms) between the plugin with DefaultPreemption enabled and the default scheduler for runs with one priority (mean ± std).}
\label{tab:latency-defpreempt1-prio1}
\adjustbox{max width=\textwidth, max totalheight=0.9\textheight}{%
\begin{tabular}{l w{c}{5.0em} w{c}{5.0em} w{c}{5.0em} w{c}{5.0em} w{c}{5.0em} w{c}{5.0em}}
\toprule
\multirow{2}{*}{\textbf{Trigger Mode}} & \multicolumn{3}{c}{\# Nodes =\,16} & \multicolumn{3}{c}{\# Nodes =\,32} \\
\cmidrule(lr){2-4}\cmidrule(lr){5-7}
 & \llap{Inter-arrival =\,}4s & 8s & 16s & 4s & 8s & 16s \\
\midrule
Scheduling-failure (blocking) & $+2487\pm1089$ & $+2966\pm2214$ & $-679\pm2186$ & $+3100\pm1746$ & $+4283\pm2644$ & $+2156\pm6819$ \\
Scheduling-failure (non-blocking) & $+2725\pm1793$ & $+2548\pm1703$ & $-1258\pm2069$ & $+2935\pm2137$ & $+5491\pm3312$ & $+965\pm5276$ \\
Periodic (blocking), 4s interval & $-146\pm451$ & $+29\pm1353$ & $-2864\pm1375$ & $-1008\pm948$ & $-846\pm1657$ & $-4552\pm4468$ \\
Periodic (non-blocking), 4s interval & $-208\pm497$ & $+38\pm1632$ & $-3551\pm1528$ & $-1000\pm1027$ & $-707\pm1272$ & $-6048\pm3682$ \\
Periodic (blocking), 8s interval & $-59\pm571$ & $+312\pm1417$ & $-3052\pm1203$ & $-692\pm1184$ & $-401\pm1662$ & $-4817\pm3359$ \\
Periodic (non-blocking), 8s interval & $-174\pm423$ & $+32\pm1401$ & $-2777\pm1382$ & $-1109\pm850$ & $+83\pm2143$ & $-4235\pm3271$ \\
\bottomrule
\end{tabular}%
}
\end{table*}

%% file: figures/generated/tables/table_defpreempt=1_priorities=4_latency.tex
\begin{table*}[htbp]
\centering
\caption{Mean paired differences in scheduling latency (ms) between the plugin with DefaultPreemption enabled and the default scheduler for runs with 4 priorities (mean ± std). p1 is the lowest priority.}
\label{tab:latency-defpreempt1-prio4}
\adjustbox{max width=\textwidth, max totalheight=0.9\textheight}{%
%
%
}
\end{table*}

%% file: figures/generated/tables/table_defpreempt=1_priorities=1_deletions.tex
\begin{table*}[htbp]
\centering
\caption{Mean paired differences in number of pod deletions between the plugin with DefaultPreemption enabled and the default scheduler for runs with one priority (mean ± std).}
\label{tab:deletions-defpreempt1-prio1}
\adjustbox{max width=\textwidth, max totalheight=0.9\textheight}{%
\begin{tabular}{l w{c}{5.0em} w{c}{5.0em} w{c}{5.0em} w{c}{5.0em} w{c}{5.0em} w{c}{5.0em}}
\toprule
\multirow{2}{*}{\textbf{Trigger Mode}} & \multicolumn{3}{c}{\# Nodes =\,16} & \multicolumn{3}{c}{\# Nodes =\,32} \\
\cmidrule(lr){2-4}\cmidrule(lr){5-7}
 & \llap{Inter-arrival =\,}4s & 8s & 16s & 4s & 8s & 16s \\
\midrule
Scheduling-failure (blocking) & $-430.0\pm104.5$ & $-87.6\pm72.3$ & $-17.6\pm23.0$ & $-557.6\pm156.2$ & $-187.6\pm75.3$ & $-77.4\pm52.0$ \\
Scheduling-failure (non-blocking) & $-480.2\pm221.9$ & $-75.8\pm57.6$ & $-7.8\pm17.0$ & $-528.2\pm172.3$ & $-210.6\pm84.7$ & $-73.8\pm65.0$ \\
Periodic (blocking), 4s interval & $+58.0\pm55.5$ & $+83.8\pm30.9$ & $+33.0\pm20.5$ & $+84.2\pm85.6$ & $+60.6\pm34.5$ & $+16.6\pm29.0$ \\
Periodic (non-blocking), 4s interval & $+35.4\pm75.7$ & $+72.6\pm48.0$ & $+54.4\pm33.1$ & $+80.0\pm104.7$ & $+55.0\pm36.1$ & $+18.4\pm40.4$ \\
Periodic (blocking), 8s interval & $-23.4\pm77.7$ & $+31.0\pm56.8$ & $+21.8\pm25.4$ & $+19.8\pm109.6$ & $+37.6\pm46.6$ & $+11.2\pm25.9$ \\
Periodic (non-blocking), 8s interval & $+2.2\pm81.7$ & $+29.4\pm46.8$ & $+22.6\pm22.1$ & $+64.4\pm60.8$ & $+46.2\pm49.3$ & $+21.0\pm15.7$ \\
\bottomrule
\end{tabular}%
}
\end{table*}

%% file: figures/generated/tables/table_defpreempt=1_priorities=4_deletions.tex
\begin{table*}[htbp]
\centering
\caption{Mean paired differences in number of pod deletions between the plugin with DefaultPreemption enabled and the default scheduler for runs with 4 priorities (mean ± std). p1 is the lowest priority.}
\label{tab:deletions-defpreempt1-prio4}
\adjustbox{max width=\textwidth, max totalheight=0.9\textheight}{%
%
%
}
\end{table*}

%% file: figures/generated/tables/table_defpreempt=1_priorities=1_optimizer_runs.tex
\begin{table*}[htbp]
\centering
\caption{Mean paired differences in number of solver runs between the plugin with DefaultPreemption enabled and the default scheduler for runs with one priority (mean ± std).}
\label{tab:optimizer_runs-defpreempt1-prio1}
\adjustbox{max width=\textwidth, max totalheight=0.9\textheight}{%
\begin{tabular}{l w{c}{5.0em} w{c}{5.0em} w{c}{5.0em} w{c}{5.0em} w{c}{5.0em} w{c}{5.0em}}
\toprule
\multirow{2}{*}{\textbf{Trigger Mode}} & \multicolumn{3}{c}{\# Nodes =\,16} & \multicolumn{3}{c}{\# Nodes =\,32} \\
\cmidrule(lr){2-4}\cmidrule(lr){5-7}
 & \llap{Inter-arrival =\,}4s & 8s & 16s & 4s & 8s & 16s \\
\midrule
Scheduling-failure (blocking) & $218\pm10.3$ & $152\pm17.5$ & $108\pm30.0$ & $191\pm22.2$ & $147\pm20.9$ & $109\pm40.4$ \\
Scheduling-failure (non-blocking) & $226\pm18.5$ & $156\pm18.0$ & $114\pm30.4$ & $184\pm24.1$ & $149\pm30.2$ & $110\pm35.1$ \\
Periodic (blocking), 4s interval & $86\pm14.4$ & $80\pm17.2$ & $75\pm28.8$ & $101\pm19.5$ & $92\pm23.4$ & $96\pm42.8$ \\
Periodic (non-blocking), 4s interval & $87\pm15.9$ & $76\pm15.6$ & $80\pm36.0$ & $96\pm15.5$ & $95\pm22.9$ & $93\pm42.6$ \\
Periodic (blocking), 8s interval & $47\pm4.1$ & $45\pm8.8$ & $51\pm16.7$ & $65\pm11.1$ & $71\pm21.6$ & $61\pm23.8$ \\
Periodic (non-blocking), 8s interval & $47\pm9.8$ & $46\pm9.7$ & $50\pm13.6$ & $63\pm14.8$ & $76\pm26.5$ & $70\pm29.5$ \\
\bottomrule
\end{tabular}%
}
\end{table*}

%% file: figures/generated/tables/table_defpreempt=1_priorities=4_optimizer_runs.tex
\begin{table*}[htbp]
\centering
\caption{Mean paired differences in number of solver runs between the plugin with DefaultPreemption enabled and the default scheduler for runs with 4 priorities (mean ± std). p1 is the lowest priority.}
\label{tab:optimizer_runs-defpreempt1-prio4}
\adjustbox{max width=\textwidth, max totalheight=0.9\textheight}{%
\begin{tabular}{l w{c}{5.0em} w{c}{5.0em} w{c}{5.0em} w{c}{5.0em} w{c}{5.0em} w{c}{5.0em}}
\toprule
\multirow{2}{*}{\textbf{Trigger Mode}} & \multicolumn{3}{c}{\# Nodes =\,16} & \multicolumn{3}{c}{\# Nodes =\,32} \\
\cmidrule(lr){2-4}\cmidrule(lr){5-7}
 & \llap{Inter-arrival =\,}4s & 8s & 16s & 4s & 8s & 16s \\
\midrule
Scheduling-failure (blocking) & $260\pm56.7$ & $193\pm44.5$ & $123\pm35.8$ & $181\pm13.5$ & $123\pm9.5$ & $80\pm15.7$ \\
Scheduling-failure (non-blocking) & $266\pm67.2$ & $190\pm30.7$ & $118\pm29.3$ & $190\pm16.5$ & $127\pm10.0$ & $84\pm23.5$ \\
Periodic (blocking), 4s interval & $104\pm41.4$ & $95\pm23.6$ & $94\pm56.1$ & $86\pm13.7$ & $71\pm8.9$ & $57\pm21.5$ \\
Periodic (non-blocking), 4s interval & $109\pm41.9$ & $117\pm67.8$ & $78\pm35.1$ & $84\pm16.1$ & $69\pm9.3$ & $54\pm23.3$ \\
Periodic (blocking), 8s interval & $56\pm25.8$ & $65\pm26.5$ & $52\pm18.4$ & $53\pm4.8$ & $46\pm5.1$ & $46\pm19.3$ \\
Periodic (non-blocking), 8s interval & $59\pm26.3$ & $58\pm10.6$ & $58\pm30.0$ & $55\pm9.9$ & $50\pm3.3$ & $40\pm15.1$ \\
\bottomrule
\end{tabular}%
}
\end{table*}

%% file: figures/generated/tables/table_defpreempt=1_priorities=1_plan_activations.tex
\begin{table*}[htbp]
\centering
\caption{Mean paired differences in number of plan activations between the plugin with DefaultPreemption enabled and the default scheduler for runs with one priority (mean ± std).}
\label{tab:plan_activations-defpreempt1-prio1}
\adjustbox{max width=\textwidth, max totalheight=0.9\textheight}{%
\begin{tabular}{l w{c}{5.0em} w{c}{5.0em} w{c}{5.0em} w{c}{5.0em} w{c}{5.0em} w{c}{5.0em}}
\toprule
\multirow{2}{*}{\textbf{Trigger Mode}} & \multicolumn{3}{c}{\# Nodes =\,16} & \multicolumn{3}{c}{\# Nodes =\,32} \\
\cmidrule(lr){2-4}\cmidrule(lr){5-7}
 & \llap{Inter-arrival =\,}4s & 8s & 16s & 4s & 8s & 16s \\
\midrule
Scheduling-failure (blocking) & $134\pm24.4$ & $87\pm17.4$ & $56\pm13.0$ & $97\pm17.1$ & $60\pm12.5$ & $36\pm4.3$ \\
Scheduling-failure (non-blocking) & $135\pm27.4$ & $94\pm15.3$ & $64\pm16.1$ & $96\pm18.5$ & $58\pm9.9$ & $37\pm9.5$ \\
Periodic (blocking), 4s interval & $51\pm7.4$ & $48\pm5.8$ & $34\pm9.5$ & $50\pm12.6$ & $32\pm9.4$ & $21\pm6.1$ \\
Periodic (non-blocking), 4s interval & $50\pm5.1$ & $47\pm3.7$ & $39\pm11.7$ & $49\pm10.2$ & $33\pm6.7$ & $23\pm5.4$ \\
Periodic (blocking), 8s interval & $32\pm4.4$ & $29\pm2.2$ & $27\pm6.9$ & $34\pm8.1$ & $21\pm6.6$ & $18\pm3.7$ \\
Periodic (non-blocking), 8s interval & $31\pm5.0$ & $30\pm2.1$ & $27\pm3.6$ & $34\pm6.5$ & $23\pm6.3$ & $19\pm4.2$ \\
Periodic (blocking), 16s interval & $17\pm5.0$ & $19\pm3.7$ & $14\pm1.8$ & $20\pm5.4$ & $17\pm4.7$ & $15\pm4.0$ \\
Periodic (non-blocking), 16s interval & $20\pm3.0$ & $18\pm4.4$ & $19\pm2.4$ & $22\pm2.8$ & $16\pm6.5$ & $15\pm3.0$ \\
Stable-queue (blocking), 4s delay & $46\pm6.5$ & $35\pm3.6$ & $29\pm5.8$ & $32\pm4.6$ & $23\pm6.1$ & $19\pm6.2$ \\
Stable-queue (non-blocking), 4s delay & $41\pm8.1$ & $37\pm2.2$ & $33\pm11.4$ & $33\pm7.6$ & $24\pm5.0$ & $17\pm5.0$ \\
Stable-queue (blocking), 8s delay & $27\pm5.3$ & $27\pm5.3$ & $23\pm3.5$ & $24\pm7.0$ & $21\pm4.7$ & $17\pm4.1$ \\
Stable-queue (non-blocking), 8s delay & $27\pm5.9$ & $28\pm4.5$ & $25\pm4.3$ & $24\pm7.0$ & $21\pm5.1$ & $17\pm6.3$ \\
Stable-queue (blocking), 16s delay & $16\pm3.8$ & $17\pm5.0$ & $16\pm3.3$ & $16\pm7.4$ & $15\pm4.0$ & $14\pm5.1$ \\
Stable-queue (non-blocking), 16s delay & $15\pm4.2$ & $18\pm4.5$ & $16\pm3.0$ & $17\pm4.5$ & $15\pm4.5$ & $20\pm4.1$ \\
\bottomrule
\end{tabular}%
}
\end{table*}

%% file: figures/generated/tables/table_defpreempt=1_priorities=4_plan_activations.tex
\begin{table*}[htbp]
\centering
\caption{Mean paired differences in number of plan activations between the plugin with DefaultPreemption enabled and the default scheduler for runs with 4 priorities (mean ± std). p1 is the lowest priority.}
\label{tab:plan_activations-defpreempt1-prio4}
\adjustbox{max width=\textwidth, max totalheight=0.9\textheight}{%
\begin{tabular}{l w{c}{5.0em} w{c}{5.0em} w{c}{5.0em} w{c}{5.0em} w{c}{5.0em} w{c}{5.0em}}
\toprule
\multirow{2}{*}{\textbf{Trigger Mode}} & \multicolumn{3}{c}{\# Nodes =\,16} & \multicolumn{3}{c}{\# Nodes =\,32} \\
\cmidrule(lr){2-4}\cmidrule(lr){5-7}
 & \llap{Inter-arrival =\,}4s & 8s & 16s & 4s & 8s & 16s \\
\midrule
Scheduling-failure (blocking) & $168\pm39.4$ & $116\pm25.0$ & $72\pm16.3$ & $175\pm14.1$ & $77\pm11.3$ & $45\pm5.3$ \\
Scheduling-failure (non-blocking) & $175\pm42.5$ & $120\pm23.5$ & $66\pm18.6$ & $138\pm12.8$ & $77\pm15.0$ & $46\pm8.2$ \\
Periodic (blocking), 4s interval & $72\pm22.3$ & $61\pm6.8$ & $41\pm11.2$ & $67\pm10.0$ & $44\pm7.0$ & $32\pm3.8$ \\
Periodic (non-blocking), 4s interval & $74\pm17.0$ & $59\pm15.3$ & $40\pm12.6$ & $63\pm9.3$ & $40\pm7.8$ & $31\pm8.5$ \\
Periodic (blocking), 8s interval & $43\pm18.4$ & $40\pm10.1$ & $32\pm7.0$ & $44\pm2.9$ & $29\pm7.8$ & $24\pm8.0$ \\
Periodic (non-blocking), 8s interval & $42\pm16.9$ & $44\pm13.0$ & $30\pm10.1$ & $46\pm7.4$ & $30\pm5.5$ & $25\pm8.5$ \\
\bottomrule
\end{tabular}%
}
\end{table*}

%% file: figures/generated/tables/table_defpreempt=0_priorities=1_usage.tex
\begin{table*}[htbp]
\centering
\caption{Mean paired differences in effective resource usage (\%) between the plugin with DefaultPreemption disabled and the default scheduler for runs with one priority (mean ± std).}
\label{tab:usage-defpreempt0-prio1}
\adjustbox{max width=\textwidth, max totalheight=0.9\textheight}{%
\begin{tabular}{l w{c}{5.0em} w{c}{5.0em} w{c}{5.0em} w{c}{5.0em} w{c}{5.0em} w{c}{5.0em}}
\toprule
\multirow{2}{*}{\textbf{Trigger Mode}} & \multicolumn{3}{c}{\# Nodes =\,16} & \multicolumn{3}{c}{\# Nodes =\,32} \\
\cmidrule(lr){2-4}\cmidrule(lr){5-7}
 & \llap{Inter-arrival =\,}4s & 8s & 16s & 4s & 8s & 16s \\
\midrule
Scheduling-failure (blocking) & $+1.2\pm0.7$ & $+1.9\pm1.1$ & $+3.0\pm0.8$ & $+1.4\pm0.6$ & $+1.5\pm1.5$ & $+2.7\pm1.3$ \\
Scheduling-failure (non-blocking) & $+1.2\pm0.8$ & $+1.9\pm1.1$ & $+3.0\pm0.7$ & $+1.2\pm0.4$ & $+1.4\pm1.2$ & $+2.5\pm1.2$ \\
Periodic (blocking), 4s interval & $+2.0\pm0.6$ & $+2.2\pm1.0$ & $+3.1\pm0.8$ & $+1.5\pm0.4$ & $+1.9\pm1.2$ & $+2.4\pm1.1$ \\
Periodic (non-blocking), 4s interval & $+1.8\pm0.6$ & $+2.2\pm1.0$ & $+3.1\pm0.8$ & $+1.7\pm0.6$ & $+1.8\pm0.9$ & $+2.3\pm1.3$ \\
Periodic (blocking), 8s interval & $+1.5\pm0.8$ & $+2.1\pm1.1$ & $+3.0\pm0.9$ & $+1.6\pm0.6$ & $+1.7\pm1.4$ & $+2.4\pm1.2$ \\
Periodic (non-blocking), 8s interval & $+1.6\pm0.8$ & $+2.1\pm1.0$ & $+3.0\pm0.8$ & $+1.6\pm0.6$ & $+1.8\pm1.0$ & $+2.6\pm1.3$ \\
\bottomrule
\end{tabular}%
}
\end{table*}

%% file: figures/generated/tables/table_defpreempt=0_priorities=4_usage.tex
\begin{table*}[htbp]
\centering
\caption{Mean paired differences in effective resource usage (\%) between the plugin with DefaultPreemption disabled and the default scheduler for runs with 4 priorities (mean ± std). p1 is the lowest priority.}
\label{tab:usage-defpreempt0-prio4}
\adjustbox{max width=\textwidth, max totalheight=0.9\textheight}{%
\begin{tabular}{l w{c}{5.0em} w{c}{5.0em} w{c}{5.0em} w{c}{5.0em} w{c}{5.0em} w{c}{5.0em}}
\toprule
\multirow{2}{*}{\textbf{Trigger Mode}} & \multicolumn{3}{c}{\# Nodes =\,16} & \multicolumn{3}{c}{\# Nodes =\,32} \\
\cmidrule(lr){2-4}\cmidrule(lr){5-7}
 & \llap{Inter-arrival =\,}4s & 8s & 16s & 4s & 8s & 16s \\
\midrule
Scheduling-failure (blocking) & $+0.4\pm1.0$ & $+1.2\pm1.6$ & $+1.4\pm1.1$ & $+0.7\pm0.8$ & $+0.7\pm0.3$ & $+0.9\pm0.7$ \\
Scheduling-failure (non-blocking) & $+0.4\pm1.0$ & $+1.1\pm1.3$ & $+1.3\pm1.2$ & $+0.7\pm0.7$ & $+0.6\pm0.3$ & $+0.9\pm0.7$ \\
Periodic (blocking), 4s interval & $+1.0\pm0.8$ & $+1.4\pm1.2$ & $+1.6\pm1.1$ & $+0.9\pm0.8$ & $+0.8\pm0.3$ & $+1.0\pm0.6$ \\
Periodic (non-blocking), 4s interval & $+1.1\pm0.7$ & $+1.5\pm1.0$ & $+1.5\pm1.0$ & $+0.9\pm0.7$ & $+0.7\pm0.3$ & $+0.9\pm0.6$ \\
Periodic (blocking), 8s interval & $+0.8\pm0.8$ & $+1.3\pm1.1$ & $+1.5\pm1.1$ & $+0.9\pm0.7$ & $+0.5\pm0.4$ & $+0.8\pm0.7$ \\
Periodic (non-blocking), 8s interval & $+0.8\pm0.8$ & $+1.4\pm1.1$ & $+1.4\pm1.1$ & $+0.8\pm0.8$ & $+0.7\pm0.2$ & $+0.9\pm0.6$ \\
\bottomrule
\end{tabular}%
}
\end{table*}

%% file: figures/generated/tables/table_defpreempt=0_priorities=1_latency.tex
\begin{table*}[htbp]
\centering
\caption{Mean paired differences in scheduling latency (ms) between the plugin with DefaultPreemption disabled and the default scheduler for runs with one priority (mean ± std).}
\label{tab:latency-defpreempt0-prio1}
\adjustbox{max width=\textwidth, max totalheight=0.9\textheight}{%
\begin{tabular}{l w{c}{5.0em} w{c}{5.0em} w{c}{5.0em} w{c}{5.0em} w{c}{5.0em} w{c}{5.0em}}
\toprule
\multirow{2}{*}{\textbf{Trigger Mode}} & \multicolumn{3}{c}{\# Nodes =\,16} & \multicolumn{3}{c}{\# Nodes =\,32} \\
\cmidrule(lr){2-4}\cmidrule(lr){5-7}
 & \llap{Inter-arrival =\,}4s & 8s & 16s & 4s & 8s & 16s \\
\midrule
Scheduling-failure (blocking) & $+2349\pm992$ & $+2954\pm1988$ & $-865\pm1938$ & $+2904\pm1329$ & $+5572\pm3599$ & $-51\pm5367$ \\
Scheduling-failure (non-blocking) & $+2499\pm1056$ & $+2832\pm1825$ & $-1013\pm1854$ & $+2762\pm1748$ & $+6153\pm4018$ & $+2771\pm6931$ \\
Periodic (blocking), 4s interval & $-196\pm471$ & $-159\pm1250$ & $-3955\pm1838$ & $-988\pm1193$ & $-1180\pm722$ & $-4839\pm4214$ \\
Periodic (non-blocking), 4s interval & $-140\pm553$ & $-51\pm1241$ & $-3304\pm1710$ & $-1075\pm1003$ & $-1185\pm1083$ & $-3896\pm3365$ \\
Periodic (blocking), 8s interval & $+21\pm551$ & $+302\pm1423$ & $-2679\pm1742$ & $-889\pm989$ & $-613\pm1804$ & $-4658\pm4628$ \\
Periodic (non-blocking), 8s interval & $+19\pm563$ & $+29\pm1053$ & $-3438\pm1709$ & $-1076\pm987$ & $-567\pm1377$ & $-5216\pm3605$ \\
\bottomrule
\end{tabular}%
}
\end{table*}

%% file: figures/generated/tables/table_defpreempt=0_priorities=4_latency.tex
\begin{table*}[htbp]
\centering
\caption{Mean paired differences in scheduling latency (ms) between the plugin with DefaultPreemption disabled and the default scheduler for runs with 4 priorities (mean ± std). p1 is the lowest priority.}
\label{tab:latency-defpreempt0-prio4}
\adjustbox{max width=\textwidth, max totalheight=0.9\textheight}{%
%
%
}
\end{table*}

%% file: figures/generated/tables/table_defpreempt=0_priorities=1_deletions.tex
\begin{table*}[htbp]
\centering
\caption{Mean paired differences in number of pod deletions between the plugin with DefaultPreemption disabled and the default scheduler for runs with one priority (mean ± std).}
\label{tab:deletions-defpreempt0-prio1}
\adjustbox{max width=\textwidth, max totalheight=0.9\textheight}{%
\begin{tabular}{l w{c}{5.0em} w{c}{5.0em} w{c}{5.0em} w{c}{5.0em} w{c}{5.0em} w{c}{5.0em}}
\toprule
\multirow{2}{*}{\textbf{Trigger Mode}} & \multicolumn{3}{c}{\# Nodes =\,16} & \multicolumn{3}{c}{\# Nodes =\,32} \\
\cmidrule(lr){2-4}\cmidrule(lr){5-7}
 & \llap{Inter-arrival =\,}4s & 8s & 16s & 4s & 8s & 16s \\
\midrule
Scheduling-failure (blocking) & $-420.4\pm95.7$ & $-103.4\pm56.2$ & $-0.8\pm30.3$ & $-549.6\pm121.6$ & $-213.0\pm71.6$ & $-66.6\pm55.0$ \\
Scheduling-failure (non-blocking) & $-401.2\pm84.5$ & $-102.4\pm73.8$ & $-0.8\pm28.1$ & $-518.8\pm151.1$ & $-202.8\pm97.7$ & $-86.6\pm57.4$ \\
Periodic (blocking), 4s interval & $+53.8\pm68.0$ & $+69.0\pm38.5$ & $+60.2\pm22.7$ & $+96.4\pm116.3$ & $+60.2\pm24.3$ & $+30.6\pm24.0$ \\
Periodic (non-blocking), 4s interval & $+43.0\pm61.9$ & $+85.0\pm28.2$ & $+42.8\pm22.7$ & $+101.2\pm78.6$ & $+64.4\pm31.9$ & $+32.8\pm20.2$ \\
Periodic (blocking), 8s interval & $-37.2\pm97.9$ & $+15.4\pm58.3$ & $+13.2\pm19.7$ & $+58.4\pm92.5$ & $+36.6\pm29.7$ & $+17.6\pm21.1$ \\
Periodic (non-blocking), 8s interval & $-12.8\pm66.8$ & $+21.0\pm40.0$ & $+18.8\pm23.4$ & $+69.4\pm61.7$ & $+34.2\pm36.1$ & $+6.8\pm34.6$ \\
\bottomrule
\end{tabular}%
}
\end{table*}

%% file: figures/generated/tables/table_defpreempt=0_priorities=4_deletions.tex
\begin{table*}[htbp]
\centering
\caption{Mean paired differences in number of pod deletions between the plugin with DefaultPreemption disabled and the default scheduler for runs with 4 priorities (mean ± std). p1 is the lowest priority.}
\label{tab:deletions-defpreempt0-prio4}
\adjustbox{max width=\textwidth, max totalheight=0.9\textheight}{%
%
%
}
\end{table*}

%% file: figures/generated/tables/table_defpreempt=0_priorities=1_optimizer_runs.tex
\begin{table*}[htbp]
\centering
\caption{Mean paired differences in number of solver runs between the plugin with DefaultPreemption disabled and the default scheduler for runs with one priority (mean ± std).}
\label{tab:optimizer_runs-defpreempt0-prio1}
\adjustbox{max width=\textwidth, max totalheight=0.9\textheight}{%
\begin{tabular}{l w{c}{5.0em} w{c}{5.0em} w{c}{5.0em} w{c}{5.0em} w{c}{5.0em} w{c}{5.0em}}
\toprule
\multirow{2}{*}{\textbf{Trigger Mode}} & \multicolumn{3}{c}{\# Nodes =\,16} & \multicolumn{3}{c}{\# Nodes =\,32} \\
\cmidrule(lr){2-4}\cmidrule(lr){5-7}
 & \llap{Inter-arrival =\,}4s & 8s & 16s & 4s & 8s & 16s \\
\midrule
Scheduling-failure (blocking) & $218\pm9.9$ & $154\pm17.3$ & $111\pm30.7$ & $195\pm26.0$ & $145\pm24.8$ & $102\pm30.6$ \\
Scheduling-failure (non-blocking) & $223\pm11.7$ & $152\pm19.7$ & $111\pm27.3$ & $182\pm19.2$ & $149\pm33.8$ & $111\pm40.2$ \\
Periodic (blocking), 4s interval & $84\pm17.6$ & $77\pm15.9$ & $76\pm26.4$ & $100\pm20.4$ & $93\pm23.0$ & $90\pm42.2$ \\
Periodic (non-blocking), 4s interval & $86\pm13.8$ & $80\pm15.6$ & $72\pm24.3$ & $96\pm18.0$ & $99\pm31.2$ & $93\pm39.2$ \\
Periodic (blocking), 8s interval & $46\pm8.7$ & $46\pm10.2$ & $52\pm14.8$ & $57\pm9.9$ & $68\pm16.3$ & $66\pm30.5$ \\
Periodic (non-blocking), 8s interval & $47\pm5.6$ & $45\pm8.9$ & $49\pm12.9$ & $60\pm11.7$ & $68\pm19.0$ & $63\pm23.0$ \\
\bottomrule
\end{tabular}%
}
\end{table*}

%% file: figures/generated/tables/table_defpreempt=0_priorities=4_optimizer_runs.tex
\begin{table*}[htbp]
\centering
\caption{Mean paired differences in number of solver runs between the plugin with DefaultPreemption disabled and the default scheduler for runs with 4 priorities (mean ± std). p1 is the lowest priority.}
\label{tab:optimizer_runs-defpreempt0-prio4}
\adjustbox{max width=\textwidth, max totalheight=0.9\textheight}{%
\begin{tabular}{l w{c}{5.0em} w{c}{5.0em} w{c}{5.0em} w{c}{5.0em} w{c}{5.0em} w{c}{5.0em}}
\toprule
\multirow{2}{*}{\textbf{Trigger Mode}} & \multicolumn{3}{c}{\# Nodes =\,16} & \multicolumn{3}{c}{\# Nodes =\,32} \\
\cmidrule(lr){2-4}\cmidrule(lr){5-7}
 & \llap{Inter-arrival =\,}4s & 8s & 16s & 4s & 8s & 16s \\
\midrule
Scheduling-failure (blocking) & $268\pm63.3$ & $189\pm28.3$ & $117\pm18.6$ & $191\pm12.5$ & $129\pm7.5$ & $84\pm20.5$ \\
Scheduling-failure (non-blocking) & $273\pm64.2$ & $180\pm32.7$ & $117\pm21.4$ & $186\pm12.2$ & $127\pm6.3$ & $86\pm20.7$ \\
Periodic (blocking), 4s interval & $107\pm31.9$ & $90\pm15.4$ & $73\pm25.3$ & $81\pm15.5$ & $76\pm4.0$ & $53\pm17.2$ \\
Periodic (non-blocking), 4s interval & $98\pm16.0$ & $87\pm14.2$ & $85\pm39.7$ & $85\pm9.8$ & $73\pm11.7$ & $60\pm18.6$ \\
Periodic (blocking), 8s interval & $51\pm6.1$ & $53\pm12.2$ & $46\pm8.8$ & $53\pm9.8$ & $56\pm4.4$ & $46\pm18.4$ \\
Periodic (non-blocking), 8s interval & $48\pm10.2$ & $57\pm20.8$ & $47\pm11.5$ & $55\pm7.9$ & $51\pm4.1$ & $45\pm14.8$ \\
\bottomrule
\end{tabular}%
}
\end{table*}

%% file: figures/generated/tables/table_defpreempt=0_priorities=1_plan_activations.tex
\begin{table*}[htbp]
\centering
\caption{Mean paired differences in number of plan activations between the plugin with DefaultPreemption disabled and the default scheduler for runs with one priority (mean ± std).}
\label{tab:plan_activations-defpreempt0-prio1}
\adjustbox{max width=\textwidth, max totalheight=0.9\textheight}{%
\begin{tabular}{l w{c}{5.0em} w{c}{5.0em} w{c}{5.0em} w{c}{5.0em} w{c}{5.0em} w{c}{5.0em}}
\toprule
\multirow{2}{*}{\textbf{Trigger Mode}} & \multicolumn{3}{c}{\# Nodes =\,16} & \multicolumn{3}{c}{\# Nodes =\,32} \\
\cmidrule(lr){2-4}\cmidrule(lr){5-7}
 & \llap{Inter-arrival =\,}4s & 8s & 16s & 4s & 8s & 16s \\
\midrule
Scheduling-failure (blocking) & $136\pm22.7$ & $91\pm18.4$ & $62\pm16.9$ & $106\pm12.0$ & $55\pm12.2$ & $37\pm4.6$ \\
Scheduling-failure (non-blocking) & $140\pm31.8$ & $89\pm19.9$ & $60\pm16.1$ & $95\pm18.6$ & $54\pm11.2$ & $37\pm6.0$ \\
Periodic (blocking), 4s interval & $53\pm2.1$ & $45\pm7.6$ & $37\pm9.9$ & $52\pm11.9$ & $35\pm4.6$ & $23\pm6.5$ \\
Periodic (non-blocking), 4s interval & $50\pm6.3$ & $47\pm2.8$ & $35\pm9.5$ & $52\pm11.7$ & $34\pm5.4$ & $22\pm6.6$ \\
Periodic (blocking), 8s interval & $30\pm6.2$ & $28\pm2.2$ & $27\pm4.6$ & $35\pm5.5$ & $24\pm8.5$ & $19\pm6.0$ \\
Periodic (non-blocking), 8s interval & $33\pm8.2$ & $28\pm2.7$ & $26\pm4.3$ & $36\pm11.1$ & $23\pm4.8$ & $20\pm4.4$ \\
Periodic (blocking), 16s interval & $18\pm1.7$ & $18\pm4.9$ & $17\pm4.0$ & $22\pm5.6$ & $16\pm4.2$ & $14\pm3.3$ \\
Periodic (non-blocking), 16s interval & $17\pm5.1$ & $18\pm2.9$ & $15\pm1.6$ & $21\pm4.9$ & $17\pm5.4$ & $14\pm4.3$ \\
Stable-queue (blocking), 4s delay & $45\pm5.0$ & $38\pm2.4$ & $32\pm4.9$ & $32\pm7.2$ & $26\pm4.1$ & $19\pm3.1$ \\
Stable-queue (non-blocking), 4s delay & $41\pm5.2$ & $38\pm2.0$ & $32\pm7.9$ & $36\pm8.4$ & $23\pm6.8$ & $20\pm5.8$ \\
Stable-queue (blocking), 8s delay & $29\pm5.5$ & $24\pm5.2$ & $26\pm4.3$ & $24\pm5.7$ & $18\pm6.9$ & $16\pm6.4$ \\
Stable-queue (non-blocking), 8s delay & $28\pm4.4$ & $29\pm5.2$ & $25\pm5.5$ & $26\pm4.7$ & $20\pm6.1$ & $14\pm4.5$ \\
Stable-queue (blocking), 16s delay & $18\pm4.5$ & $18\pm5.3$ & $16\pm1.9$ & $16\pm5.7$ & $15\pm5.1$ & $13\pm2.7$ \\
Stable-queue (non-blocking), 16s delay & $16\pm2.7$ & $17\pm4.2$ & $15\pm4.1$ & $17\pm4.4$ & $14\pm3.2$ & $12\pm2.6$ \\
\bottomrule
\end{tabular}%
}
\end{table*}

%% file: figures/generated/tables/table_defpreempt=0_priorities=4_plan_activations.tex
\begin{table*}[htbp]
\centering
\caption{Mean paired differences in number of plan activations between the plugin with DefaultPreemption disabled and the default scheduler for runs with 4 priorities (mean ± std). p1 is the lowest priority.}
\label{tab:plan_activations-defpreempt0-prio4}
\adjustbox{max width=\textwidth, max totalheight=0.9\textheight}{%
\begin{tabular}{l w{c}{5.0em} w{c}{5.0em} w{c}{5.0em} w{c}{5.0em} w{c}{5.0em} w{c}{5.0em}}
\toprule
\multirow{2}{*}{\textbf{Trigger Mode}} & \multicolumn{3}{c}{\# Nodes =\,16} & \multicolumn{3}{c}{\# Nodes =\,32} \\
\cmidrule(lr){2-4}\cmidrule(lr){5-7}
 & \llap{Inter-arrival =\,}4s & 8s & 16s & 4s & 8s & 16s \\
\midrule
Scheduling-failure (blocking) & $169\pm41.8$ & $116\pm26.3$ & $67\pm16.8$ & $139\pm6.3$ & $79\pm12.2$ & $48\pm6.8$ \\
Scheduling-failure (non-blocking) & $173\pm44.2$ & $114\pm23.9$ & $68\pm13.5$ & $135\pm5.2$ & $74\pm13.9$ & $48\pm10.5$ \\
Periodic (blocking), 4s interval & $69\pm14.4$ & $54\pm11.1$ & $39\pm7.3$ & $64\pm11.3$ & $41\pm8.1$ & $30\pm4.6$ \\
Periodic (non-blocking), 4s interval & $68\pm11.6$ & $53\pm7.1$ & $39\pm7.0$ & $65\pm6.3$ & $41\pm8.6$ & $32\pm3.2$ \\
Periodic (blocking), 8s interval & $36\pm10.6$ & $36\pm10.6$ & $26\pm3.8$ & $45\pm5.4$ & $32\pm4.9$ & $23\pm2.6$ \\
Periodic (non-blocking), 8s interval & $35\pm11.5$ & $35\pm9.6$ & $29\pm7.3$ & $44\pm3.5$ & $32\pm6.6$ & $24\pm2.1$ \\
Periodic (blocking), 16s interval & $18\pm5.9$ & $20\pm5.6$ & $18\pm3.5$ & $27\pm3.5$ & $22\pm6.5$ & $18\pm3.1$ \\
Periodic (non-blocking), 16s interval & $18\pm3.7$ & $20\pm5.1$ & $20\pm3.9$ & $25\pm3.2$ & $23\pm7.6$ & $18\pm4.7$ \\
Stable-queue (blocking), 4s delay & $53\pm10.8$ & $51\pm9.0$ & $37\pm8.0$ & $46\pm6.3$ & $31\pm7.9$ & $27\pm3.7$ \\
Stable-queue (non-blocking), 4s delay & $55\pm10.5$ & $49\pm7.6$ & $33\pm8.7$ & $48\pm3.7$ & $34\pm5.8$ & $25\pm3.6$ \\
Stable-queue (blocking), 8s delay & $35\pm7.5$ & $34\pm6.2$ & $28\pm4.3$ & $38\pm3.7$ & $27\pm4.1$ & $22\pm2.6$ \\
Stable-queue (non-blocking), 8s delay & $34\pm9.0$ & $32\pm6.6$ & $27\pm4.5$ & $35\pm3.3$ & $28\pm7.7$ & $21\pm2.6$ \\
Stable-queue (blocking), 16s delay & $18\pm4.4$ & $19\pm3.4$ & $19\pm2.1$ & $21\pm4.5$ & $22\pm4.8$ & $17\pm2.4$ \\
Stable-queue (non-blocking), 16s delay & $17\pm4.6$ & $21\pm5.8$ & $20\pm3.4$ & $23\pm4.0$ & $22\pm3.3$ & $20\pm7.0$ \\
\bottomrule
\end{tabular}%
}
\end{table*}